\documentclass[fleqn,usenatbib]{mnras}

\usepackage{newtxtext,newtxmath}

\usepackage[T1]{fontenc}

\DeclareRobustCommand{\VAN}[3]{#2}
\let\VANthebibliography\thebibliography
\def\thebibliography{\DeclareRobustCommand{\VAN}[3]{##3}\VANthebibliography}

\usepackage{graphicx}	
\usepackage{amsmath}	

\title[Standardizing microlensed Ia SN]{How to Break the Mass Sheet Degeneracy with the Lightcurves of Microlensed Type Ia Supernovae}

\author[Weisenbach et al.]{Luke Weisenbach,$^{1,}$\thanks{E-mail: weisluke@alum.mit.edu}
Thomas Collett,$^{1}$
Ana Sainz de Murieta,$^{1}$
Coleman Krawczyk,$^{1}$
\newauthor
Georgios Vernardos,$^{2,3}$ Wolfgang Enzi,$^{1}$ Andrew Lundgren$^{1}$
\\
{$^{1}$Institute of Cosmology and Gravitation, University of Portsmouth, Burnaby Road, Portsmouth, PO1 3FX, UK}\\
$^{2}$Department of Physics and Astronomy, Lehman College of the City University of New York, Bronx, NY 10468, USA\\
$^{3}$Department of Astrophysics, American Museum of Natural History, Central Park West and 79th Street, NY 10024, USA
} 

\date{Accepted XXX. Received YYY; in original form ZZZ}

\pubyear{2023}

\begin{document}
\label{firstpage}
\pagerange{\pageref{firstpage}--\pageref{lastpage}}
\maketitle

\begin{abstract}
The standardizable nature of gravitationally lensed Type Ia supernovae (glSNe Ia) makes them an attractive target for time delay cosmography, since a source with known luminosity breaks the mass sheet degeneracy. It is known that microlensing by stars in the lensing galaxy can add significant stochastic uncertainty to the unlensed luminosity which is often much larger than the intrinsic scatter of the Ia population. In this work, we show how the temporal microlensing variations as the supernova disc expands can be used to improve the standardization of glSNe Ia. We find that SNe are standardizable if they do not cross caustics as they expand. We estimate that this will be the case for $\approx$6 doubly imaged systems and $\approx$0.3 quadruply imaged systems per year in LSST. At the end of the ten year LSST survey, these systems should enable us to test for systematics in $H_0$ due to the mass sheet degeneracy at the $1.00^{+0.07}_{-0.06}$\% level, or $1.8\pm0.2$\% if we can only extract time delays from the third of systems with counter images brighter than $i=24$ mag. 
\end{abstract}




\section{Introduction}
\label{sec:intro}

Gravitational lensing is a powerful probe for understanding astrophysics and cosmology. Lensing is of particular use for constraining the expansion history of the Universe, since it is a geometric probe of the angular diameter distances between observer, lens and source \citep{1991ApJ...378L...5N}. When a time variable source is multiply imaged by a gravitational lens, the time delays between images are inversely proportional to the Hubble constant $H_0$ \citep{2016A&ARv..24...11T}. Some measurements of lensed quasars \citep{2020MNRAS.498.1420W} have found $H_0 = 73.3^{+1.7}_{-1.8}$ km s$^{-1}$ Mpc$^{-1}$, in agreement with local measurements from Type Ia supernovae \citep{2022ApJ...934L...7R} but at odds with CMB measurements from Planck \citep{2020A&A...641A...6P}. However, measurements are dependent on the mass distribution of the lens. Relaxing some assumptions and including non-lensing constraints has also lead to results which are consistent with those of Planck \citep{2020A&A...643A.165B}. While gravitational lensing has the potential to direct some light on the existing tension between $H_0$ measurements, a thorough understanding of systematics and degeneracies of lens modelling is necessary.

Historically time delays have been measured from lensed quasars \citep{2020A&A...640A.105M}, though supernovae were originally proposed \citep{1964MNRAS.128..307R}. Lensed supernovae (glSNe) offer several benefits over quasars: they evolve over much shorter timescales requiring months of monitoring as opposed to years, and they fade away enabling follow-up observations of the lensed host. Type Ia glSNe are even better: they have well understood lightcurves, and are `standardizable' candles, with a scatter of $\approx$ 0.15 mags \citep{2014AJ....147..118R, 2022ApJ...938..110B}. The fact that Type Ia supernovae have a known intrinsic luminosity means that their magnification can be directly measured, helping to break lens modelling degeneracies.

In order to infer the value of $H_0$ from a lensed time variable source, one needs to know the mass distribution of the lensing galaxy. The problem is that lensing observables alone cannot uniquely tell us about that mass distribution. A lensing model that predicts the positions, relative magnifications, and time delays between observed images can be rescaled 
with a sheet of mass in such a way that none of the observables change except for the time delays \citep{1985ApJ...289L...1F, 2014A&A...564A.103S}. This mass sheet degeneracy (MSD) must be broken to extract the Hubble constant from time delay measurements, and this is one of the main systematic uncertainties in modern time delay cosmography \citep{2020A&A...643A.165B}. The MSD is broken if you know the absolute magnification of the background source, which is possible with a lensed standard candle such as a glSNe Ia. 
 
Unfortunately, there is one main barrier to using glSNe Ia to break the MSD: microlensing by stars. The image we observe is split into unresolvable microimages that are (de)magnified by the stars in the lensing galaxy \citep{1981ApJ...244..756Y, 1986ApJ...301..503P, 2024SSRv..220...14V}. We can only observe the total flux of these unresolved images, which can be (de)magnified from the macro-model. We need to know the macrolensing magnification to break the MSD, but the presence of microlensing adds significant stochastic noise \citep{2006ApJ...653.1391D}. In the worst case scenario this can introduce up to a magnitude of scatter \citep{2021ApJ...922...70W}, which would make microlensed glSNe Ia no longer standardizable candles \citep{2017arXiv171107919Y}.

\cite{2018MNRAS.478.5081F} were the first to examine if some lensed images might suffer sufficiently small amounts of microlensing that they remain standardizable. By analyzing the microlensing of a uniform disk, they found that regions of low magnification and low stellar density have microlensing scatter comparable to the intrinsic scatter in the luminosity of an unlensed SN Ia\footnote{Throughout this paper we will refer to a glSNe Ia as standardizable if the microlensing scatter is comparable to or less than the intrinsic scatter in the luminosity of an unlensed SN Ia (0.15 mags). The microlensing scatter is defined as the half-width of the inner 68\% of the microlensing magnification probability distribution. Microlensing magnification distributions are non-Gaussian and skewed with a high-magnification tail \citep[see, e.g., Figure 4 of ][]{2013MNRAS.434..832V}, hence this definition.}. However, they did not use temporal information in their inference.
As the supernova expands, it averages over different length scales of microlensing fluctuations. The glSNe never becomes sufficiently large to completely average out to the macro-model magnification, but the shape of the lightcurve can teach us about the nature of the microimages and hence inform us about the amplitude of the absolute microlensing magnification.

In what follows, we provide a discussion on the theoretical reasoning behind why certain regions of microlensing parameter space are or are not standardizable in Section \ref{sec:why_standardizable}. We discuss simulations for a simple model in Section \ref{sec:sims_and_data}. We discuss in Section \ref{sec:lightcurve_selection} various methods and implementations for either selecting microlensing lightcurves that are standardizable or inferring a posterior of the microlensing magnification given an observed lightcurve. We extend from our single point to the entirety of useful microlensing parameter space in Section \ref{sec:covering_parameter_space}. In Section \ref{sec:estimating_num_standardizable} we provide an estimate on the number of standardizable glSNe Ia expected to be discovered in the next decade from a forecasted LSST population, while discussing some of the limitations of our work in Section \ref{sec:limitations}. We provide our conclusions in Section \ref{sec:conclusions}.

\section{Standardizable regions of microlensing parameter space}
\label{sec:why_standardizable}

\begin{figure*}
	\includegraphics[width=2\columnwidth]{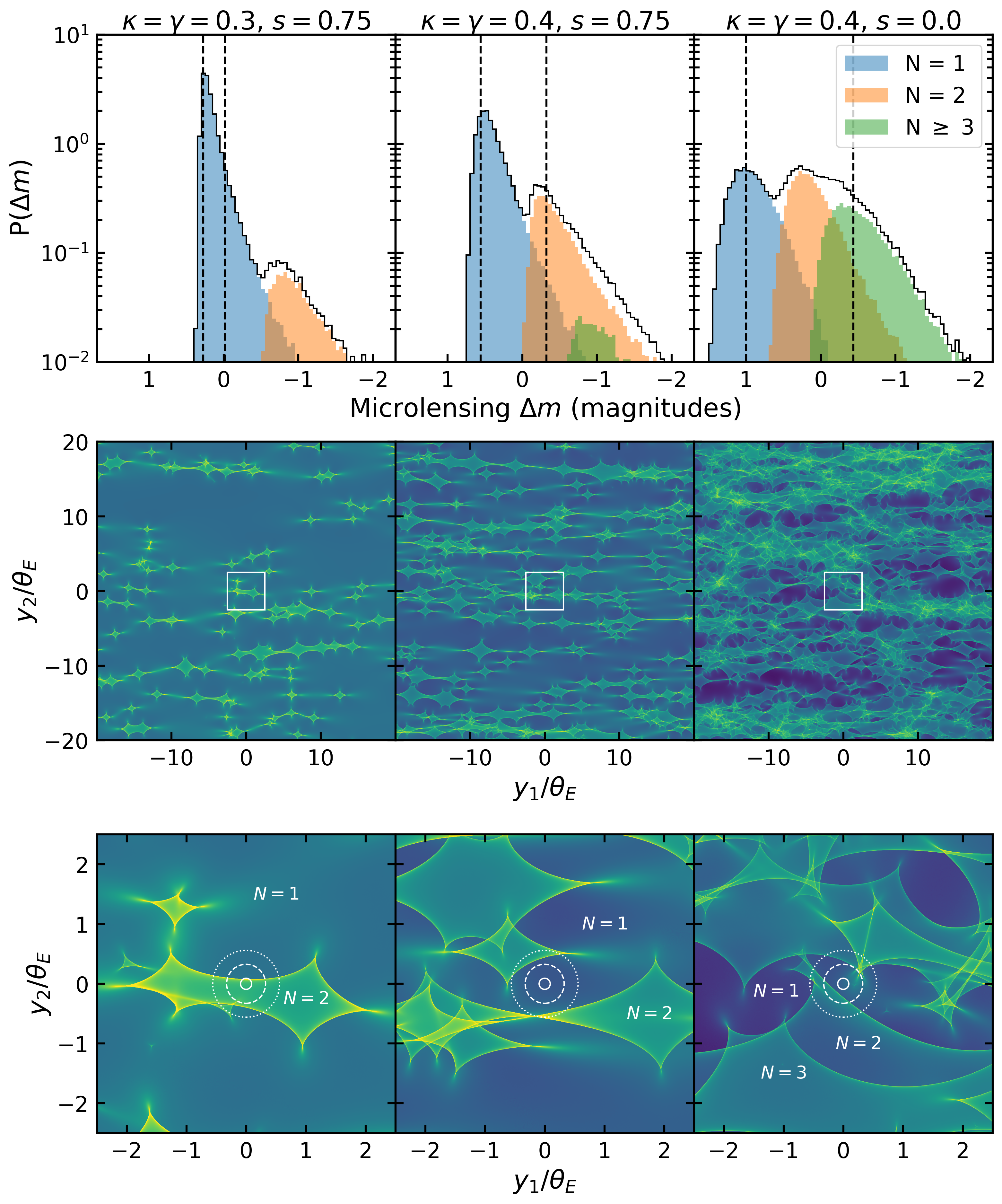}
    \caption{Top: Microlensing histograms for systems that are (from left to right) standardizable, intermediate, and unstandardizable. The vertical dashed lines mark the inner 68\% of the histogram. The subhistograms are for those regions completely outside the caustics ($N=1$), inside one caustic ($N=2$), or deeper in the caustic network ($N\geq 3$). Middle: Sample caustic patterns that produce the microlensing histograms. Bottom: Zoom of the caustics for the indicated central square of the larger maps. The solid, dashed, and dotted circles denote the size of a fiducial supernovae 0, 50, and 100 days after peak luminosity respectively.}
    \label{fig:hists_caustics_zoom}
\end{figure*}

In this section, we investigate why the time evolution of a microlensed lightcurve should be sensitive to the absolute microlensing magnification, and hence inform us about which glSNe images are standardizable. This section focuses on the theoretical underpinnings of why it should be possible to select a subset of standardizable microlensed images, while Section \ref{sec:lightcurve_selection} will show how to do so with realistic observables.

\subsection{The importance of the number of microimages}
\label{subsec:where_and_why}

Microlensing fluctuations are highly dependent on where the image forms relative to the lensing galaxy. There are  three main parameters of importance: the lens macro-model convergence and shear at the macroimage location, $\kappa$ and $\gamma$, and the ratio of stellar to total matter density at the location of the image, $1 - s$ (where $s$ is the smooth matter fraction, which is predominantly the dark matter fraction but also includes any other smooth baryonic components). 

\cite{2018MNRAS.478.5081F} and \cite{2021ApJ...922...70W} have already shown that microlensed systems with low macro-magnification and low stellar density are standardizable; see for example Figure 5 of \cite{2024SSRv..220...13S}. The physical reason that systems with low magnification and stellar density are standardizable is that the caustic network is \textit{sparse} -- the majority of the source plane is dominated by regions where the source does not lie within any microcaustics. Fundamentally, this is related to the idea that the mean number of microminima of the time delay surface \citep{1992A&A...258..591W, 2003ApJ...583..575G} for such systems is small. This is opposed to regions of parameter space with higher magnification or higher stellar density where the caustic network is more dense \citep[see, e.g., Figure 12 of][]{2024SSRv..220...14V}, raising the expected number of microimages. Regions of the source plane can be indexed by how many caustics a source lies within, which we will denote with $N=1,2,3,...$\footnote{More precisely, $N$ denotes the number of microminima \citep{2003ApJ...583..575G}. $N\geq 1$ for macrominima. Macrosaddles have an additional $N=0$ region; see Appendix \ref{app:macrosaddles}.}.

Figure \ref{fig:hists_caustics_zoom} shows three microlensing systems: one that would be standardizable with the \cite{2018MNRAS.478.5081F} method, an intermediate scenario, and one that is completely unstandardizable. Each of the microlensing histograms is bimodal, and an examination of the magnification maps and caustic structures reveals why: the histogram can be decomposed into subhistograms for each value of $N$ \citep{1992ApJ...386...30R, 2003ApJ...583..575G, 2011MNRAS.411.1671S}. These subhistograms are offset from each other, since increasing $N$ by 1 means the creation of another pair of microimages which introduces a jump in the minimum allowable magnification. The relative importance of the subhistograms is set by the density of the caustic network.

\begin{enumerate}
    \item For the standardizable system, the majority of the source plane consists of regions with $N=1$, with a small fraction consisting of $N=2$. The scatter is dominated by the $N=1$ region, so the width of the microlensing histogram is small.\\
    \item For the intermediate system, the relative fraction of $N=1$ is lower and $N=2$ is higher, making the bimodality more prominent and increasing the scatter. \\
    \item The completely unstandardizable system has approximately equal fractions for $N=1$ and $N=2$. Since the subhistograms are offset, the total microlensing histogram is very broad.
\end{enumerate}

The statement that a microlensed system is standardizable is really a statement that the likelihood of lying inside a microcaustic is low. If we could determine the value of $N$, many glSNe would be standardizable. Unfortunately resolving the microimages is currently impossible. But there is a key piece of data that \textit{is} observable, namely the lightcurve of the expanding supernova. The temporal variations as the supernova expands contain information about where it lies within the caustic network.

\subsection{Constraining the number of microimages}
\label{subsec:when}

\begin{figure}
	\includegraphics[width=\columnwidth]{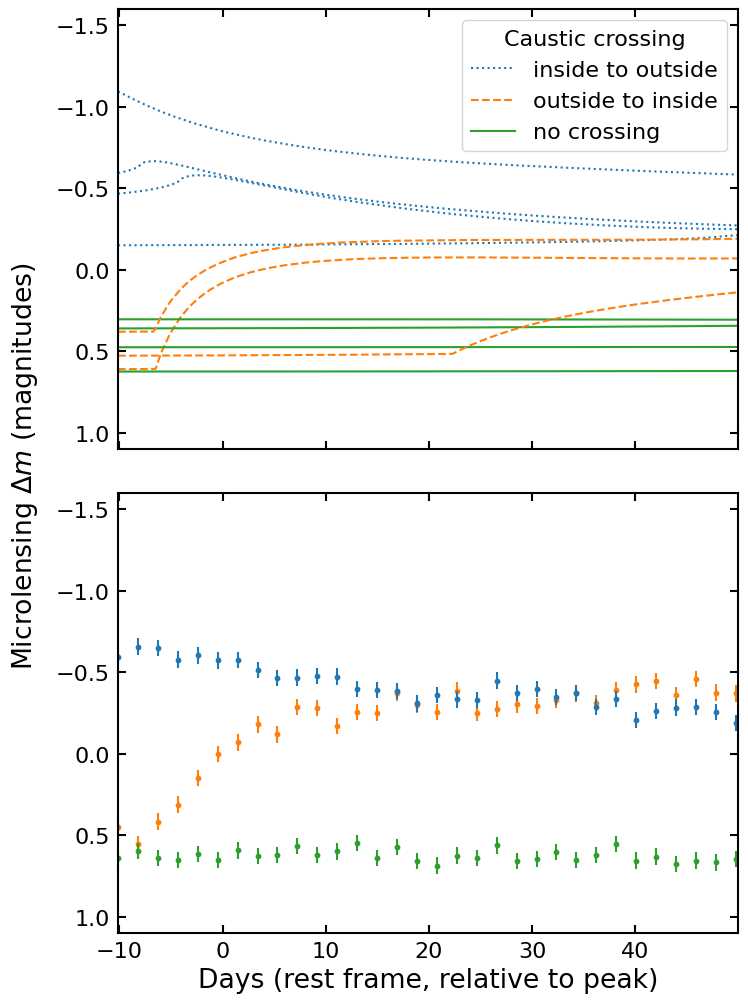}
    \caption{Top: Sample microlensing lightcurves for a source that i) crosses a caustic inside to outside (blue), ii) crosses a caustic outside to inside (orange), or iii) does not cross a caustic (green), within the time period examined. Bottom: Sample noisy lightcurves, assuming a 2 day cadence in the source rest frame and 0.05 mag Gaussian noise.}
    \label{fig:lightcurves_scenarios}
\end{figure}

Since supernovae expand overtime, a source that lies deep in the microcaustic network must eventually grow to the point that it crosses a caustic. The exact timescales involved and the rate of change of the lightcurve depend on the expansion velocity, the sizes of the caustics, and the mean spacing of the caustics; the latter two depend on the mass of the microlenses, $\kappa$, $\gamma$, and $s$. Looking at Figure \ref{fig:hists_caustics_zoom}, it is fairly easy to convince oneself that sources starting in regions with $N\gtrsim 3$ will typically expand to cross a caustic more quickly than sources located elsewhere. Since caustics represent lines of extremely high magnification, caustic crossings induce substantial changes in the microlensed supernova lightcurve. There are three basic lightcurve scenarios possible as the source expands which we illustrate in Figure \ref{fig:lightcurves_scenarios}: 

\begin{enumerate}
    \item The source lies \textit{near} and \textit{inside} (on the more magnified side of) one of the microcaustics. The micro magnification rises as the expansion approaches the caustic and then falls when parts of the disc fall outside of the microcaustic.
    \item The source lies \textit{near} and \textit{outside} (on the less magnified side of) one of the microcaustics. The micro magnification sharply rises as parts of the disk cross the caustic.
    \item The source lies \textit{sufficiently far} from the microcaustics for the timescale of interest. The micro magnification is broadly insensitive to the size of the SN disk. 
\end{enumerate} There can certainly be more complicated behaviour, for example, if the source lies closer to a cusp or overlapping caustics, but these are much rarer.


\begin{table}
	\centering
	\caption{Microlensing magnification scatter for an expanding source which has not crossed a caustic up to some time for the intermediate configuration ($\kappa = \gamma = 0.4, s = 0.75$). Values were found by convolving one large scale map. Fractions indicate what percentage of locations on the source plane have not crossed a caustic by the indicated time. The first row gives results for all of the source plane under consideration, while subsequent rows show the decomposition into regions with various values of $N$.}
	\label{tab:large_map_scatter}
	\begin{tabular}{lcccccc} 
		\hline
		& \multicolumn{2}{c}{Point source} & \multicolumn{2}{c}{Peak} & \multicolumn{2}{c}{50 days past peak} \\
		\hline
		 & Fraction & Scatter & Fraction & Scatter & Fraction & Scatter\\
		All & 1 & 0.420 & 0.840 & 0.335 & 0.564 & 0.191\\
		\hline
        $N=1$ & 0.799 & 0.187 & 0.722 & 0.178 & 0.539 & 0.167\\
        $N=2$ & 0.186 & 0.338 & 0.112 & 0.170 & 0.025 & 0.126\\
        $N\geq 3$ & 0.015 & 0.389 & 0.005 & 0.108 & - & -\\
		\hline
	\end{tabular}
\end{table} 


\begin{figure*}
	\includegraphics[width=1.8\columnwidth]{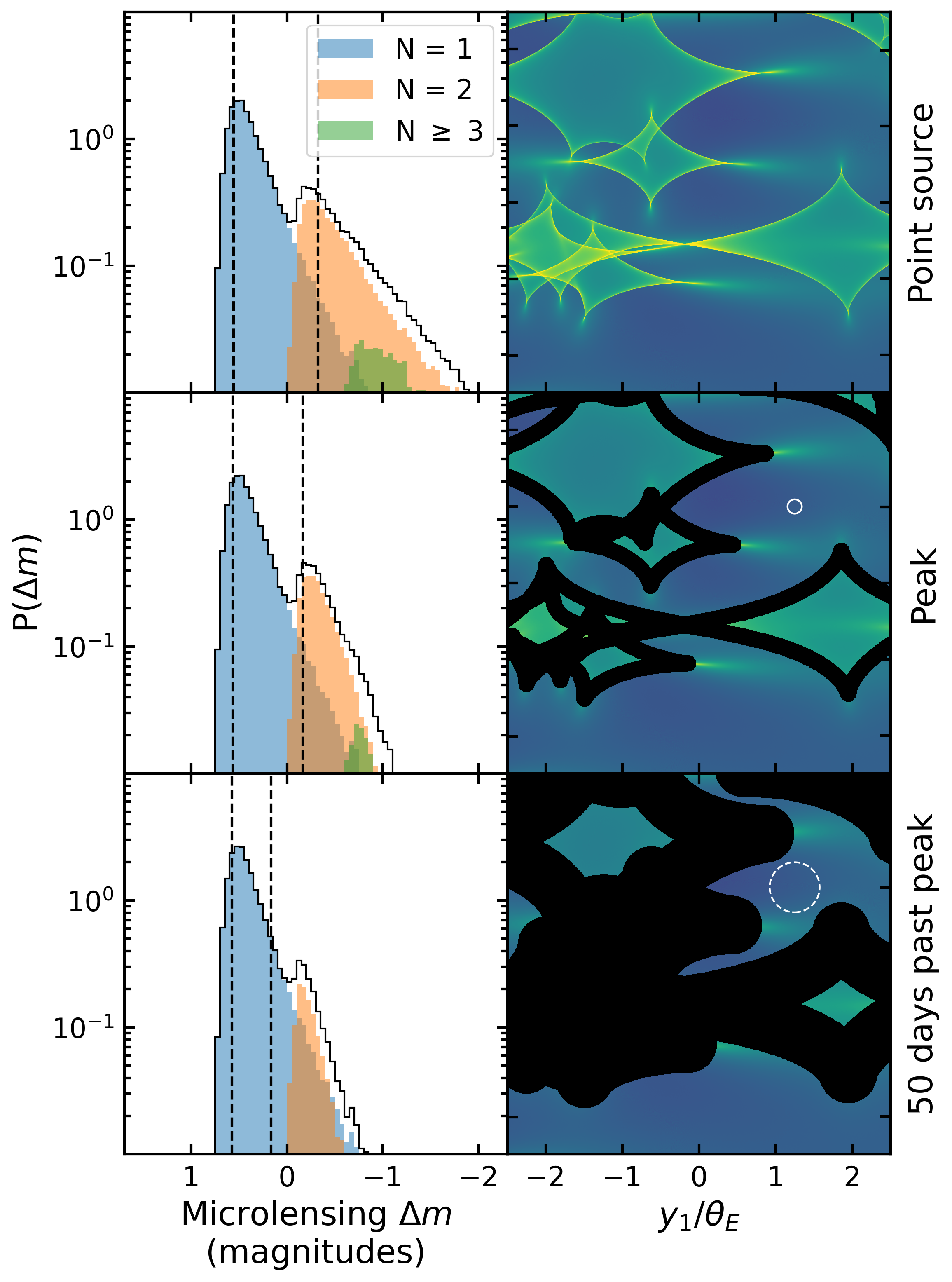}
    \caption{Left: Microlensing magnification distributions for the intermediate scenario of Figure \ref{fig:hists_caustics_zoom}. The vertical dashed lines mark the inner 68\% of the histograms. While the point source histogram is for the entire source plane, the subsequent histograms are only for those regions of the source plane where an expanding source will not cross a caustic by some given time. Right: Visualization of the ruled out regions in the source plane. The solid and dashed white circles in the magnification maps denote the size of a fiducial supernova at each time period.}
    \label{fig:hists_vs_SN_size}
\end{figure*}

The shape of the lightcurve provides information about where the source lies within the caustic network. Clearly microlensing lightcurves that are flat originate far from caustics and so are not likely to be highly magnified. One would expect that this subpopulation will show much less micromagnification scatter, and hence be more standardizable.

We can quantify this by tracking whether or not a source placed within the microlensing map for the intermediate system ($\kappa=\gamma=0.4,s=0.75$) crosses a caustic in a given time period.  A visualization of how this affects the microlensing magnification distribution is shown in Figure \ref{fig:hists_vs_SN_size}. Results are shown in Table \ref{tab:large_map_scatter}. The high magnification tail of the distribution comes from regions of the source plane that are either i) near the inside of fold caustics, ii) near the outside of cusps, or iii) deep inside the caustics. 
These regions are quickly ruled out by the absence of caustic crossings. The magnification distribution for non-crossing events shrinks and becomes much more standardizable. 

A problem with this approach is that configurations with dense caustic networks --- high macro-magnification images or images forming in regions of high stellar density --- will have very few source positions that do not cross a caustic as the SN expands. This approach also does not work for saddlepoint images as they have a low magnification tail which cannot be ruled out by the absence of a caustic crossing (see Appendix \ref{app:macrosaddles}). However, we expect intermediate systems to be standardizable under this approach and for it to be a substantial improvement upon previous analyses.

\section{Simulations and Data}
\label{sec:sims_and_data}

In this section, we discuss the assumptions used for our simulations, the simulated data that we will use throughout the paper, and sources of noise that are relevant for the lightcurves.

\subsection{Assumptions}

We will use a simple model for a microlensed supernova: we take the supernova to be a uniformly luminous disk that is expanding in radius constantly with time. While more complicated supernova models such as those used by \cite{2018ApJ...855...22G} and \cite{2019A&A...631A.161H} may technically be more accurate, \cite{2005ApJ...628..594M} and \cite{2019MNRAS.486.1944V} have shown that microlensing fluctuations are most sensitive to the half-light radius of the source as opposed to its luminosity profile. We furthermore use the following assumptions throughout this work: 

\begin{enumerate}
    \item Our lens is at a redshift $z=0.5$ and our source is at a redshift $z=0.8$, in a flat $\Lambda$CDM universe using the parameters of \cite{2020A&A...641A...6P}. Results loosely depend on the redshifts.
    \item The microlenses are all of mass $0.3M_\odot$, which determines the Einstein radius $\theta_E$. Conversions to other masses can be made using the fact that $\theta_E\propto\sqrt{M}$.
    \item The supernova has a constant expansion rate of $10^4$ km/s \citep{2020ApJ...895L...5P}. The main purpose of this is to convert from size to approximate times where necessary.
    \item The supernova peaks at 20 days in its rest frame. Days listed are days in the rest frame of the supernova unless otherwise specified.
    \item We will assume that we can separate the effects of microlensing from the intrinsic variations of the SN lightcurve, up to a constant uncertainty of 0.05 mag. This is comparable to the uncertainties in the model fitting of type Ia SNe \citep[e.g. SALT3,][see Figure \ref{fig:sn_template_errors}]{2021ApJ...923..265K}, which will set a fundamental floor on our ability to separate out the time evolution of the microlensing\footnote{Unless it is possible to use the multiple microlensed images of the glSNe to improve upon classical Ia template fitting.}. Figure \ref{fig:lightcurves_scenarios} shows some example noisy microlensing lightcurves.
\end{enumerate}

\subsection{Simulations}

We generate 100,000 microlensed lightcurves for the intermediate system parameters ($\kappa=\gamma=0.4$, $s=0.75$) to serve as our dataset. We use the inverse ray shooting method \citep{1986A&A...166...36K, 1990PhDT.......180W} to create magnification maps that are $\approx10\theta_E \times 10\theta_E$ and $\approx$10,000 x 10,000 pixels with a pixel scale of $0.001\theta_E$, or $\approx$ a quarter of a day of supernova expansion. We use enough stars to capture the bulk of the magnification \citep{1986ApJ...306....2K}, and distribute them in a rectangular region to reduce computational complexity while still accounting for the average microlensing deflection \citep{2022ApJ...931..114Z}. We can fit 10 x 10 = 100 expanding uniform disks onto a single map with no overlap after $\approx$ 100 days, requiring 1000 maps. We generate the maps and perform the convolutions on GPUs
. Since we are only interested in lightcurves from sources that do not overlap on the source plane (not correlated), we do not need to convolve the entirety of the maps, greatly reducing the computation time needed. In addition to creating the lightcurves for each expanding source, we use a GPU version of Hans Witt’s method \citep{1990A&A...236..311W} to find the caustics of each star field in order to track whether the expanding disk crosses any caustics throughout the entirety of the lightcurve. 


\section{Selecting standardizable lightcurves}
\label{sec:lightcurve_selection}

In this section, we discuss how, in practice, to infer the posterior of the microlensing magnification given an observed lightcurve. 
We start by presenting two simple criteria for picking out lightcurves which do not cross caustics and discuss some of the difficulties that might arise due to noise when using these criteria. We then discuss how a bank of lightcurves can be used to estimate the amount of microlensing (de)magnification. Next, we examine how neural network regression can be used to predict the microlensing (de)magnification. Finally, we use a neural network to classify the lightcurves into two categories: did or did not cross a caustic. 

Throughout this section, we assume glSNe lightcurves are observed from 5 days before peak up to 50 days after peak in the SN rest frame, with a 2 day cadence in the rest frame.

\subsection{Simple criteria}
\begin{figure}
	\includegraphics[width=\columnwidth]{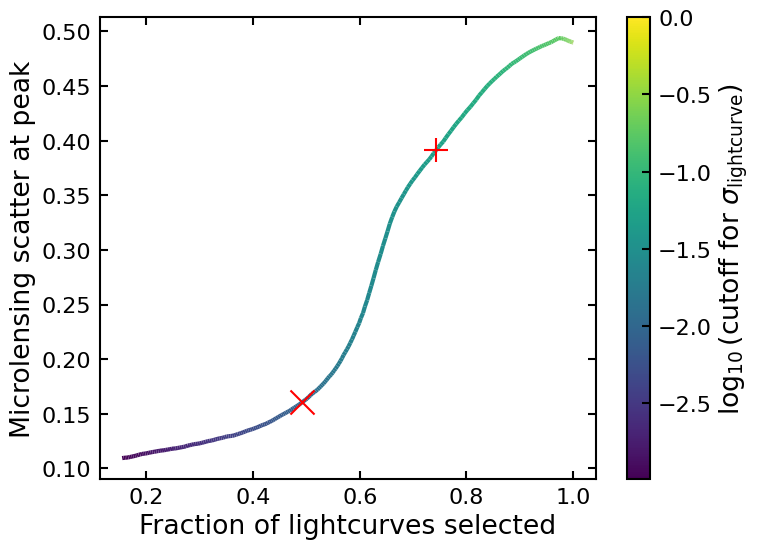}
    \caption{Given the standard deviation $\sigma_{\text{lightcurve}}$ of the lightcurve data points, we can select a subset (fraction) of lightcurves that have $\sigma_{\text{lightcurve}}$ less than some desired cutoff. Lowering the cutoff reduces the fraction of lightcurves selected and their scatter. The red ``+'' marks where the cutoff for  $\sigma_{\text{lightcurve}}=0.05$, while the red ``x'' marks where the cutoff for $\sigma_{\text{lightcurve}}=0.01$.}
    \label{fig:std_dev_selection}
\end{figure}

We consider perhaps the simplest metric: measuring the standard deviation of our simulated lightcurves $\sigma_{\text{lightcurve}}$. We can then determine the fraction of simulated lightcurves that have $\sigma_{\text{lightcurve}}$ less than some cutoff value, and what the microlensing scatter at peak is for those lightcurves. Results are shown in Figure \ref{fig:std_dev_selection}. By selecting lightcurves which have small standard deviations, we select a subset with low scatter. A given amount of noise in the data sets a lower limit for the cutoff however, limiting the utility of this metric. In practice this noise is likely to come from imperfect knowledge of the unlensed SN lightcurve, rather than observational noise. If the standard deviation can only be recovered with 0.05 mag precision the improvement is marginal - 70\% of microlensing lightcurves will be consistent with flat and only the most extreme caustic crossings can be excluded. If a precision of 0.01 mag is achievable (which is likely the case if SALT3 mismatches correlate with time) then half of the lightcurves will be consistent with flat and the standardizability for this half of the dataset improves to approximately 0.15 magnitudes.

\subsection{A bank of lightcurves}
Since we are able to simulate microlensing curves, the statistically rigorous way to infer microlensing magnifications is to compare observations with similar simulated data \citep{2004ApJ...605...58K}.

We assume that an observed lightcurve can be parameterized by a single value, the amount of microlensing (de)magnification at peak, $\mu$. We then have that \begin{equation}
    P(\mu | D) \propto P(D | \mu) P(\mu)
\end{equation} There is additionally a nuisance parameter, the position $\mathbf{y}$ of the source within the microcaustics, that must be marginalized over. The marginalization is approximated as a sum \begin{equation}
    \label{eq:p_mu_given_d}
    P(\mu | D) \approx \sum_{\mathbf{y}} P\left(D | \mu(\mathbf{y})\right)  P(\mathbf{y})
\end{equation} over a finite number of source positions, i.e. a finite number of simulated lightcurves. The likelihood $P\left(D | \mu(\mathbf{y})\right)$ is given by a $\chi^2$ statistic of the difference between the data and the simulations. Thus, $P(\mu)$ is determined by summing up the finite collection (bank) of lightcurves, where each lightcurve in the bank is weighted by how well its shape matches the observed lightcurve.

We separate our 100,000 lightcurves into two sets: 90,000 lightcurves with no noise to serve as perfect members of the bank, and 10,000 lightcurves with noise to serve as our mock observed data. The lightcurves are all shifted in magnitudes such that their means are each 0. This way, there is only relative knowledge of how their shapes evolve over time. 

\begin{figure}
    \centering
	\includegraphics[width=\columnwidth]{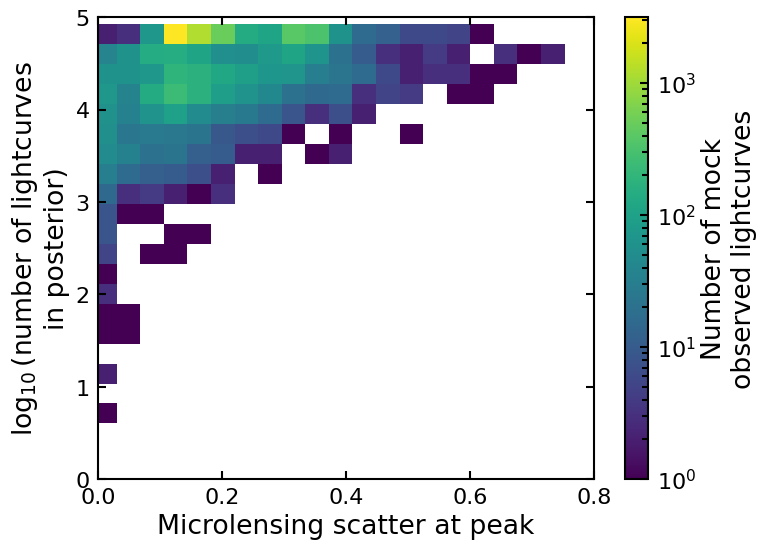}
    \caption{Inferred microlensing scatter at peak luminosity for our 10,000 mock observed lightcurves, versus the number of bank lightcurves that make up the inner 95\% of their posteriors. The majority of mock observed lightcurves have scatter less than 0.2 mags -- 33\% fall in the yellow bin, with a scatter of $\approx$0.12 mags. However, some of the lightcurves are matched only by a handful of bank lightcurves and so their posteriors are likely spurious. }
    \label{fig:bank_posteriors}
\end{figure}

We use Equation \ref{eq:p_mu_given_d} to calculate the microlensing magnification posterior for each of the 10,000 mock observed lightcurves. We then calculate the scatter of the posterior for each lightcurve. The majority of mock observed lightcurves have microlensing scatters less than 0.2 mags. This is roughly a factor of 2 improvement from the point source microlensing histogram, indicating that information about the lightcurve shape can narrow down the predicted microlensing (de)magnification. Figure \ref{fig:bank_posteriors} shows these scatters as a function of how many bank lightcurves are similar.

The majority of the bank and the dataset consists of lightcurves with no distinguishing features (flat) and these are easily standardized as seen in Figure \ref{fig:bank_posteriors}.  In contrast, there are few lightcurves which have greater time variability, i.e. caustic crossings. The lack of similar curves in the bank makes the inferred posteriors unreliable: the approximation in Equation \ref{eq:p_mu_given_d} only works if the bank is sufficiently well sampled. One could simulate more lightcurves to give a reliable posterior for everything \citep[see, e.g.,][where the order of $10^8$ lightcurves are used]{2004ApJ...605...58K}, but since the bulk of the lightcurves are flat, and these are standardizable, we leave this for future work. 

\subsection{Machine learning -- regression}

We train a neural network to predict the (de)magnification due to microlensing. The network is a fully connected network with 2 hidden layers -- simple, but sufficient for our purposes. The size of the input layer is determined by the observation length and cadence of the lightcurve, while the two hidden layers each have half as many neurons as the input layer to avoid overfitting. We take our set of 100,000 noisy lightcurves and set aside 80,000 as as training data, 10,000 as validation data, and 10,000 as test data. The label for each lightcurve in the training and validation sets is the amount of (de)magnification at peak supernova luminosity. Training is stopped when the training loss (mean squared error) on the validation data stops decreasing. The network is then applied to the 10,000 test lightcurves. 

\begin{figure}
    \centering
	\includegraphics[width=\columnwidth]{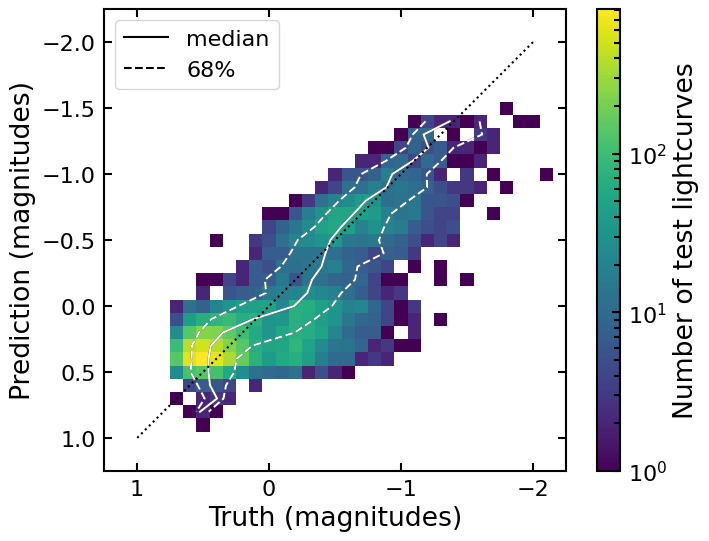}
    \caption{Predicted versus true microlensing (de)magnification from the regression neural network. The white solid and dashed lines mark the median and inner 68\% of the distribution of true values for lightcurves within the same bin of predictions. The black dotted line marks where prediction = truth.}
    \label{fig:nn_prediction_vs_truth}
\end{figure}

Figure \ref{fig:nn_prediction_vs_truth} shows the predicted microlensing magnifications compared to the true values. The peak in the distribution at 0.5 mag is due to all of the flat lines. The neural network has learned i) the average magnification of the flat lines (which make up the bulk of the data), and ii) that it can can minimize the error by assigning the mean value of the flat $N=1$ lines to all of them. There are, however, flat lines that come from the $N=2$ region which therefore have an incorrect prediction. This is the reason for a slight tail in the distribution to the right of the peak, which cannot be removed with the simple point estimator of this network. More complicated networks that return a full posterior should be able to resolve this problem \citep[e.g. Bayesian Neural Networks,][]{2022BNN}.

The neural network made some progress on the remaining data, which show a roughly even amount of scatter around the predicted = truth line with widths of 0.2-0.25 magnitudes. This is slightly greater than the assumed intrinsic supernova scatter of 0.15 magnitudes. Furthermore, the number of lightcurves with bright predicted magnifications is small, making it somewhat difficult to be as confident in their scatter. While the small number of lightcurves with bright magnifications is due to the particular point we picked in microlensing parameter space, and could potentially be remedied with more simulations, the improvements to the flat lightcurves are in line with Table \ref{tab:large_map_scatter} and our expectations. 

\subsection{Machine learning -- classification}

\begin{figure}
    \centering
	\includegraphics[width=\columnwidth]{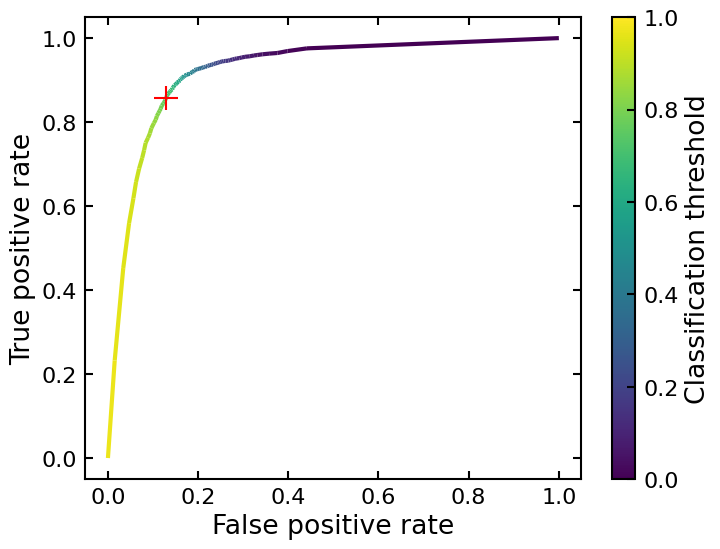}\\
	\includegraphics[width=\columnwidth]{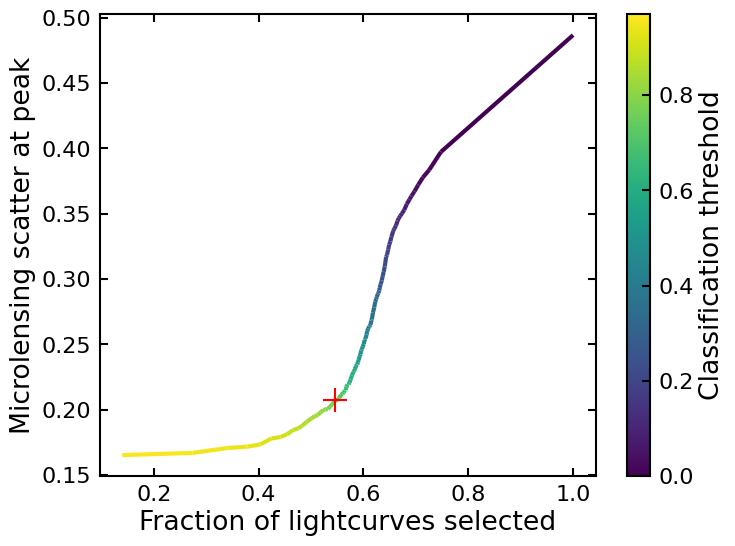}
    \caption{Top: ROC curve for the classification neural network. Bottom: Fraction of lightcurves selected as not crossing a caustic, and their scatter, as a function of classification threshold. The red plus symbol marks the 75\% classification threshold in both figures.}
    \label{fig:nn_categorization}
\end{figure}

We train a different neural network to classify the lightcurves into two categories: did it cross a caustic or not? We use the same simple network architecture as before but instead of training the network to learn the underlying magnification at peak, the training set is tagged 1 if it did not cross a caustic and 0 if it did cross. The output of the network then is no longer the magnification at peak, but the probability of a lightcurve belonging to either of the two categories. Figure \ref{fig:nn_categorization} shows the ROC curve for the network when applied to the test data, along with the fraction of lightcurves selected as a function of the classification threshold and their associated scatter. At a classification threshold of 75\% (roughly the threshold with the highest true positive rate and lowest false positive rate), slightly less than 60\% of the lightcurves are inferred not to have crossed a caustic, and the scatter from those lightcurves is roughly 0.2 mags -- consistent with the expectations from Table \ref{tab:large_map_scatter}. 

\subsection{Remarks}

We have shown that there are a variety of viable methods for reducing the scatter for the intermediate scenario microlensing parameters that we chose ($\kappa = \gamma = 0.4, s = 0.75$). Ultimately, each method gives approximately the same results, with the same underlying physical reason: flat microlensing lightcurves, which come from supernovae that do not cross a caustic as they expand over a long enough timescale, are more standardizable. The performance of the lightcurve bank and the regression neural network yield some further improvement for lightcurves that do cross a caustic, but since these events start off with much higher variance the improvement will not have much impact on the inference of the Hubble constant from a population of glSNe. The simple method of using the standard deviation of the lightcurve is sufficient to quickly select out the standardizable microlensing lightcurves.


\section{Covering microlensing parameter space}
\label{sec:covering_parameter_space}

\begin{figure*}
    \centering
	\includegraphics[width=2\columnwidth]{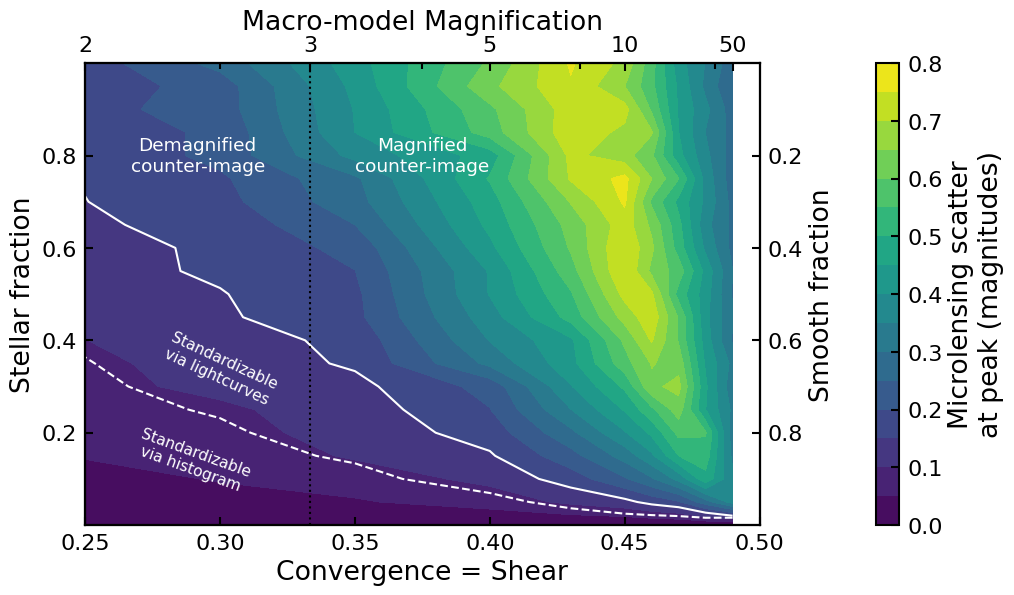}
    \caption{Contour plot showing the scatter over microlensing parameter space for those lightcurves which the neural network predicts did not show any caustic crossings. The lightcurves used assume we observe every 2 days from 5 days before peak up to 50 days after peak (in the rest frame of the supernova) and have 0.05 mag noise on the data. The solid white line denotes the contour of 0.15 mag microlensing scatter for the lightcurves. The dashed white line denotes the contour of 0.15 mag scatter when considering only the point source histograms, i.e. no time-series information \citep{2018MNRAS.478.5081F}. The vertical black dotted line marks the boundary where the counter-image of a singular isothermal sphere is demagnified (left) or magnified (right).}
    \label{fig:scatter_over_parameter_space}
\end{figure*}

Whilst the previous section focused on a particular set of parameters ($\kappa = \gamma = 0.4, s = 0.75$), we expect that other regions of microlensing parameter space with lower magnifications or lower stellar fractions can be similarly improved. We turn now to covering the rest of the useful parameter space. We opt to use the final neural network-based approach where lightcurves are sorted into two categories.

Any triplet of model values $(\kappa, \gamma, s)$ can be transformed into a doublet of $(\kappa=\gamma, s)$ \citep{1986ApJ...301..503P, 2004ApJ...605...58K, 2014ApJS..211...16V, 2014ApJ...793...96S}, which is applicable for a singular isothermal elliptical potential. We therefore cover the space from $\kappa=\gamma=0.25$\footnote{Strong lensing does not occur below $\kappa=0.25$ in an isothermal sphere, although it will if the system is more elliptical.} to $\kappa=\gamma=0.5$ and from $s=0$ to $s=1$. For each point sampled in this space, we create 5,000 microlensing lightcurves using the same procedures discussed in Section \ref{sec:sims_and_data}. 

We take 4000 lightcurves from every point sampled in the parameter space to create our training set and use 1000 lightcurves from each point as validation to avoid overfitting\footnote{The final amount taken from each sample is technically smaller, as we want equal number of lightcurves which do or do not cross a caustic to train on. We ultimately end up with $\approx$300,000 lightcurves in our training set and $\approx$75,000 as validation.}. This is done for two reasons: first, we would expect to get a fair number of flat lightcurves in the standardizable regions and non-flat lightcurves in the unstandardizable regions; second, by training the network on a sample of lightcurves that come from everywhere in the space and therefore show potential complexities from multiple caustic crossing events, we expect it to be more general and applicable. 

Once the training is completed, we generate 5000 lightcurves at each point in parameter space to test the performance of the network. We calculate the scatter on those lightcurves which the network says does not cross a caustic. Figure \ref{fig:scatter_over_parameter_space} shows the results. The neural network selects only a fraction of the lightcurves (shown in Figure \ref{fig:fraction_selected}). However, the standardizable region of parameter space has been improved when compared to considering just the point-source histogram \citep[i.e. using no time-series information from the lightcurve,][]{2018MNRAS.478.5081F}. More systems with magnified counter-images are standardizable, and a large fraction of parameter space with demagnified counter images are now standardizable as well.

\section{Standardizable LSST lensed Ia supernovae}
\label{sec:estimating_num_standardizable}

The discussion up to this point has focused on the theoretical improvements that could be made to standardizing microlensed Ia supernovae based on temporal information from their lightcurves. In this section, we discuss the practical difficulties of actually observing a glSN Ia which can be standardized. We then estimate the number of standardizable glSNe Ia to be discovered by LSST in the next decade. Finally, we examine how well we can constrain systematics in measurements of $H_0$ from the mass sheet degeneracy with a sample of standardizable glSNe Ia. 

Given that the unstandardizable saddlepoint macroimages are the trailing images, standardizing lensed supernovae heavily relies on discovering the leading image (or two, if the system is a quad) before peak. For a doubly imaged system, there is only one chance to standardize it. For a quad, there can be a second chance, but only for the rare quads with two standardizable images discovered early enough. 

\subsection{Generating (micro-)lensed supernovae samples}
We follow \cite{2023MNRAS.526.4296S} to generate a population of glSNe Ia and determine the fraction that will be observable, useful for time-delay cosmography, and standardizable. While \cite{2023MNRAS.526.4296S} focused on unresolved lightcurves, the methods can be used to create a catalogue of a large population of glSNe Ia observed in the $i$ band. 

Since we need to measure the microlensing in each image, the images need to be resolvable; we limit ourselves to systems where the minimum image separation is $0.8''$. In order to be a good candidate for time delay cosmography, the supernova images also need to have an appreciable ($\gtrsim$ 10 days) time delay, as time delays can be recovered to within a day or two \citep{2021ApJ...908..190P, 2022A&A...658A.157H}. We additionally impose that the system be discovered no later than 10 days before the peak of the first (second) image for doubles (quads) in the observer frame. This allows a long baseline to measure the evolution of the microlensing signal as the SN disk grows as in Section \ref{sec:lightcurve_selection}. 

We estimate the stellar fractions at the image locations by assuming an elliptical de Vaucouleurs profile for the light \citep{2019MNRAS.483.5583V} with effective radii calculated from the scaling relation in \cite{2009MNRAS.394.1978H}. Following \citet{2006ApJ...653.1391D}, we assume isothermal total density profiles and normalise the stellar component such that the maximum stellar fraction is 1. This will make our forecasts somewhat pessimistic since observations prefer lower stellar fractions, but it depends on the stellar initial mass function \citep{2010ApJ...724..511A}.

We furthermore limit ourselves to systems where microlensing cannot demagnify the first image enough to be unobservable. This is done to ensure that any observed image would be a fair draw from the microlensing magnification distribution, rather than an unfair sampling from the brighter tail of the distribution. This would naively restrict us to redshifts where the unlensed supernova magnitude is greater than the detection limit of the survey; when accounting for microlensing, the minimum magnification is $\mu = (1 - s\cdot\kappa)^{-2}$ which, depending on the macro-model and the stellar fraction for the image, allows us to push to slightly farther redshifts. We use 24th magnitude at this minimum magnification as our detection limit, approximately the value appropriate for LSST.

\subsection{Results}

\begin{figure}
    \centering
	\includegraphics[width=\columnwidth]{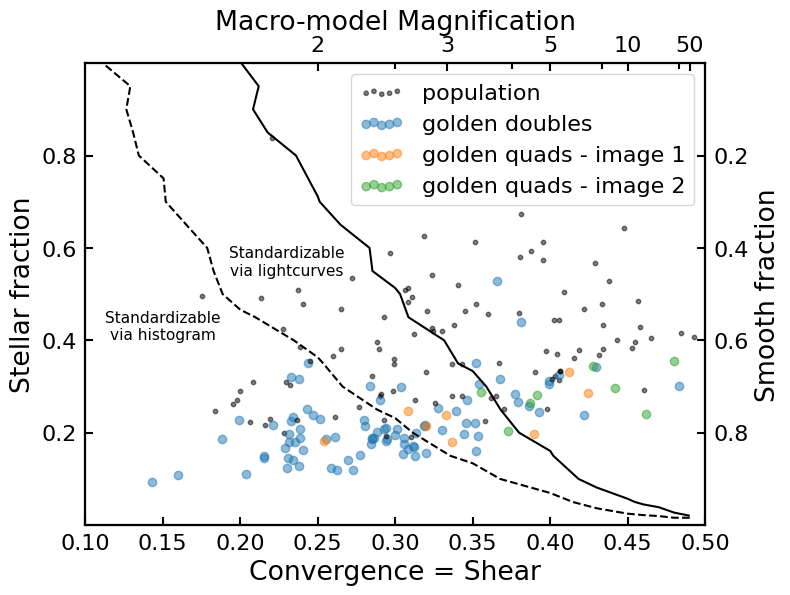}
    \caption{One realisation from the catalogue of glSNe systems expected after a decade of LSST observations. The dashed and solid black lines denote the boundaries where systems are standardizable either via the microlensing magnification histogram or their lightcurves.}
    \label{fig:sampled_systems}
\end{figure}

We can now estimate rates of standardizable glSNe and make forecasts for breaking the mass sheet degeneracy with these systems.

\subsubsection{Rate estimates}

We start with a population of glSNe from \cite{2023MNRAS.526.4296S} with a rate of 13 doubles (4 quads) per year which are discovered early enough for standardization to be possible. We show one realization from the catalogue of such systems in Figure \ref{fig:sampled_systems}. Of those, 8.3 (1.3) per year are cosmologically golden, with time delays $>10$ days and minimum image separations of $0.8''$. Of the golden systems, 6.0 doubles and 0.3 quads per year  have standardizable images with flat lightcurves. 

\subsubsection{Constraints on $H_0$}

The MSD is governed by a parameter $\lambda$. The propagation of magnification uncertainties onto $\lambda$ is given in Appendix \ref{app:prop_of_uncertainties}. If we assume the intrinsic supernova scatter of 0.15 mag dominates the uncertainty on the observed magnification and microlensing dominates the uncertainty on the model magnification, we can infer how each standardizable system will constrain the systematics from the mass sheet transformation. Combining the $\approx$60 expected cosmological golden standardizable doubles, we find that they should constrain the population average $\langle\lambda\rangle$ to $1.00^{+0.07}_{-0.06}$\% fractional uncertainty, and therefore detect systematics in $H_0$ from the MSD at the 1\% level. We note that if the intrinsic supernova scatter is lower (0.1 mag), this changes to a $0.74^{+0.06}_{-0.05}$\% fractional uncertainty on $\langle\lambda\rangle$ and $H_0$.

\subsection{Comparison to previous works}

Our results are more pessimistic than those of \cite{2018MNRAS.478.5081F}. The main factor behind this is a decrease in the estimated number of systems to be discovered with LSST; compare, e.g., \cite{2019ApJS..243....6G}, \cite{2019MNRAS.487.3342W}, and \cite{2023MNRAS.526.4296S} (particularly their discussion in Section 4.3). We additionally imposed more restrictions on what systems might be considered cosmologically useful and standardizable than \cite{2018MNRAS.478.5081F}. 

\cite{2023arXiv231204621A} estimate slightly higher rates of detection for LSST, which in part come about due to considering alternative image detection methods that do not rely solely on magnification \citep[see also, e.g.,][]{2019MNRAS.487.3342W}. However, detection methods such as image multiplicity miss early time information which could be key for standardizing microlensed lightcurves. 

Comparing our results to those ignoring the time evolution of the lightcurve  \citep[i.e., roughly following][and using just the point source microlensing histograms]{2018MNRAS.478.5081F}, we find that $\approx$20\% more of the cosmological golden systems will be standardizable. We were able to  substantially increase the size of the standardizable region of microlensing parameter space in Figure \ref{fig:scatter_over_parameter_space}, which gives a 30\% improvement on the number of standardizable systems in the whole population. However, the majority of systems with long time delays tend to come from the regions that were already standardizable under the considerations of \cite{2018MNRAS.478.5081F}.

A byproduct of our methods is a decrease in the scatter on the systems that were already standardizable, at the cost of losing a small fraction of lightcurves. However, we get similar constraints on $H_0$ for our simulated systems if we ignore information about the shape of the lightcurves: the rate of standardizable doubles decreases to 5.1 per year, which leads to a $1.11^{+0.09}_{-0.07}$\% fractional uncertainty on $H_0$ from the MSD.
This suggests that model macro-parameters are the main driving factors in standardizability. This comes from the fact that whilst we have tightened $P(\mu)$ for all macrominima, the intrinsic scatter in the Ia population sets a floor on the utility of an individual system in breaking the MSD. 

\subsection{Observing the counter-images of standardizable images}

\begin{figure}
    \centering
	\includegraphics[width=\columnwidth]{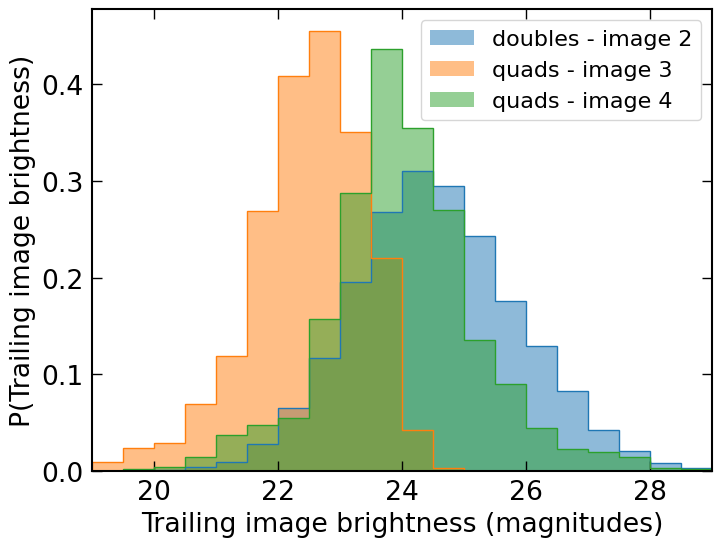}
    \caption{Distribution of counter image Sloan i band magnitudes under the lens macro-model only for the standardizable systems.}
    \label{fig:counter_image_mags}
\end{figure}

Actually doing time-delay cosmography with these systems requires us to detect the trailing saddlepoint image(s). Figure \ref{fig:counter_image_mags} shows the distribution of counter image magnifications at peak, from the lens macro-model only. The distribution peaks at 24.5 mags in the $i$ band for the doubles, 24 mags for image 4 of the quads, and 22.5 mags for image 3 of the quads. However, this ignores the fact that saddlepoint images are more susceptible to microlensing demagnifications \citep{2002ApJ...580..685S}. Followup observations of the trailing image(s) will be difficult, but possible. 

When considering the number of standardizable systems with bright counter-images, our lightcurve method represents a substantial improvement upon the previous histogram method of \citet{2018MNRAS.478.5081F}. If we consider only the doubles which have a trailing image that peaks brighter than 25th mag, our yearly rate of standardizable doubles drops from 6.0 to 3.9, compared to the 3 per year expected when ignoring lightcurve shape. After ten years, we should have 39 systems and a $1.28^{+0.12}_{-0.09}$\% fractional uncertainty on $H_0$ from the MSD. For counter-images brighter than 24th magnitude we expect 2 standardizable systems per year and $1.8 \pm 0.2$\% fractional uncertainty on $H_0$ from the MSD.

\section{Limitations}
\label{sec:limitations}

The detection of any microlensing signal is limited due to uncertainties in the supernova model. We have pessimistically assumed 0.05 mag uncertainty on each point in our lightcurves. Whilst this is comparable to the mismatch in SALT3, we have neglected temporal or chromatic correlations, both of which might be able to improve upon our assumed 0.05 mag uncertainty.

The standardizable lightcurves are typically \textit{demagnified} from the macro-model prediction. Although in our study we limited ourselves to simulated systems where microlensing could never demagnify the first image below the detection threshold of LSST, we must keep in mind that this will not always be the case. Supernovae at higher redshifts $(z\gtrsim 1)$ begin to suffer from a microlensing magnification bias in LSST, as microlensing can demagnify (relative to the macro-model) the first image of systems below the detection threshold of the survey. Incorporating such systems would require a careful simulation of their Malmquist bias. 

We have assumed that the lens macro-model parameters for the images are perfectly known. In reality, uncertainties on $\kappa$, $\gamma$, and $s$ require properly marginalizing over a region of the parameter space in Figure \ref{fig:scatter_over_parameter_space}, rather than taking the scatter at an individual point. If uncertainties on the macro-model parameters are not large, the slow variations of the scatter in the standardizable region suggests that there will be minimal changes to results.

While our simulated systems have values of $s$ calculated with simple assumptions, $s$ should properly be inferred from the mass to light ratio of the lensing galaxy. If we were to use a Salpeter or Chabrier IMF, the smooth matter fraction would increase \citep{2018MNRAS.478.5081F}, providing a small boost in the number of standardizable systems. A rough estimation of this effect by halving the stellar fractions used leads to little change in our results: 7.6 standardizable doubles per year and 0.4 quads, with the doubles constraining $H_0$ to $0.90^{+0.06}_{-0.05}$\% fractional uncertainty from the MSD.


Furthermore, there is an additional dependence of microlensing on the mass function of the lenses \citep{2004ApJ...613...77S} that is usually ignored. This dependence is difficult to manifest in microlensed quasars for physical reasons \citep{2006ApJ...645..835L} but the implication for lensed supernovae has not been investigated.

While we considered the simple model of a uniform expanding disc, real supernovae will have a 2D intensity profile that must be convolved with the microlensing magnification pattern \citep{2018ApJ...855...22G, 2019A&A...631A.161H}. We don't expect any changes to our conclusions due to the fact that the standardizable lightcurves have essentially constant microlensing magnification.

\section{Conclusions}
\label{sec:conclusions}

In this work, we have investigated how the temporal information from microlensing lightcurves can be used to reduce the uncertainty on the unlensed SN magnitude. We have shown that it is possible to select a subset of glSNe Ia images which can be used to break the mass sheet degeneracy. The main idea we have put forward is that by selecting lightcurves of supernovae which do not cross any microcaustics as they expand, you can restrict yourself to regions of the source plane typically outside the microcaustics which have lower amounts of scatter. Our method relies on having a long enough sequence of observations to rule out caustic crossings. 

Using a simulated sample population of lensed Type Ia supernovae with resolvable ($>0.8''$ separation) images and long ($>10$ days) time delays, we estimate the number of detectable, standardizable glSNe Ia systems to be discovered by LSST in the next decade as $\approx$60 doubly imaged systems and $\approx$3 quadruply imaged system. The doubles can constrain systematics in $H_0$ from the MSD at a $1.00^{+0.07}_{-0.06}$\% precision level. While the first image for the majority of cosmologically useful doubles should be able to break the mass sheet degeneracy, there will be observational challenges in following up the second image to measure the time delay. This is mostly driven by the faintness of the (typically demagnified) counter-images, which have a median brightness of $\approx$25 mag before accounting for additional microlensing (de)magnification. If we are only able to followup counter-images brighter than 24th magnitude, only a third of the double image systems are retained, and the fractional uncertainty on the breaking of the MSD degrades to  $1.8 \pm 0.2$\%.

Time delay measurements will already require high-quality observations up to and after peak supernova luminosity \citep{2019ApJ...876..107P, 2019A&A...631A.161H}. In that sense, no additional data should be required to determine whether the expanding supernova crosses a microcaustic or not. Given the observational cost of followup, it should be focused on systems where the first image is most likely to be standardizable. When the microlensing effect does not vary with time we essentially know the intrinsic shape of the unlensed SN lightcurve. By focusing on standardizable systems it will be easier to infer time delays, and after 10 years of LSST allow us to test the impact of the MSD on $H_0$ at $1\%$ precision.

\section*{Acknowledgements}

We would like to thank D’Arcy Kenworthy for his kind assistance in answering questions about SALT3. We would also like to thank Nikki Arendse and Matt O'Dowd for extremely helpful discussion.

Numerical computations were done on the Sciama High Performance Compute (HPC) cluster which is supported by the ICG, SEPNet, and the University of Portsmouth. 

This work has received funding from the European Research Council (ERC) under the European Union’s Horizon 2020 research and innovation programme (LensEra: grant agreement No. 945536). TC is funded by the Royal Society through a University Research Fellowship. For the purpose of open access, the authors have applied a Creative Commons Attribution (CC BY) license to any Author Accepted Manuscript version arising.

\section*{Data Availability}

The simulated data from this work can be made available upon reasonable request to the corresponding author. 



\bibliographystyle{mnras}
\bibliography{bibliography.bib} 

\begin{thebibliography}{}
\makeatletter
\relax
\def\mn@urlcharsother{\let\do\@makeother \do\$\do\&\do\#\do\^\do\_\do\%\do\~}
\def\mn@doi{\begingroup\mn@urlcharsother \@ifnextchar [ {\mn@doi@}
  {\mn@doi@[]}}
\def\mn@doi@[#1]#2{\def\@tempa{#1}\ifx\@tempa\@empty \href
  {http://dx.doi.org/#2} {doi:#2}\else \href {http://dx.doi.org/#2} {#1}\fi
  \endgroup}
\def\mn@eprint#1#2{\mn@eprint@#1:#2::\@nil}
\def\mn@eprint@arXiv#1{\href {http://arxiv.org/abs/#1} {{\tt arXiv:#1}}}
\def\mn@eprint@dblp#1{\href {http://dblp.uni-trier.de/rec/bibtex/#1.xml}
  {dblp:#1}}
\def\mn@eprint@#1:#2:#3:#4\@nil{\def\@tempa {#1}\def\@tempb {#2}\def\@tempc
  {#3}\ifx \@tempc \@empty \let \@tempc \@tempb \let \@tempb \@tempa \fi \ifx
  \@tempb \@empty \def\@tempb {arXiv}\fi \@ifundefined
  {mn@eprint@\@tempb}{\@tempb:\@tempc}{\expandafter \expandafter \csname
  mn@eprint@\@tempb\endcsname \expandafter{\@tempc}}}

\bibitem[\protect\citeauthoryear{{Arendse} et~al.,}{{Arendse}
  et~al.}{2023}]{2023arXiv231204621A}
{Arendse} N.,  et~al., 2023, \mn@doi [arXiv e-prints]
  {10.48550/arXiv.2312.04621}, \href
  {https://ui.adsabs.harvard.edu/abs/2023arXiv231204621A} {p. arXiv:2312.04621}

\bibitem[\protect\citeauthoryear{{Auger}, {Treu}, {Bolton}, {Gavazzi},
  {Koopmans}, {Marshall}, {Moustakas}  \& {Burles}}{{Auger}
  et~al.}{2010}]{2010ApJ...724..511A}
{Auger} M.~W.,  {Treu} T.,  {Bolton} A.~S.,  {Gavazzi} R.,  {Koopmans}
  L.~V.~E.,  {Marshall} P.~J.,  {Moustakas} L.~A.,   {Burles} S.,  2010,
  \mn@doi [\apj] {10.1088/0004-637X/724/1/511}, \href
  {https://ui.adsabs.harvard.edu/abs/2010ApJ...724..511A} {724, 511}

\bibitem[\protect\citeauthoryear{{Birrer} et~al.,}{{Birrer}
  et~al.}{2020}]{2020A&A...643A.165B}
{Birrer} S.,  et~al., 2020, \mn@doi [\aap] {10.1051/0004-6361/202038861}, \href
  {https://ui.adsabs.harvard.edu/abs/2020A&A...643A.165B} {643, A165}

\bibitem[\protect\citeauthoryear{{Birrer}, {Dhawan}  \& {Shajib}}{{Birrer}
  et~al.}{2022}]{2022ApJ...924....2B}
{Birrer} S.,  {Dhawan} S.,   {Shajib} A.~J.,  2022, \mn@doi [\apj]
  {10.3847/1538-4357/ac323a}, \href
  {https://ui.adsabs.harvard.edu/abs/2022ApJ...924....2B} {924, 2}

\bibitem[\protect\citeauthoryear{{Brout} et~al.,}{{Brout}
  et~al.}{2022}]{2022ApJ...938..110B}
{Brout} D.,  et~al., 2022, \mn@doi [\apj] {10.3847/1538-4357/ac8e04}, \href
  {https://ui.adsabs.harvard.edu/abs/2022ApJ...938..110B} {938, 110}

\bibitem[\protect\citeauthoryear{{Dobler} \& {Keeton}}{{Dobler} \&
  {Keeton}}{2006}]{2006ApJ...653.1391D}
{Dobler} G.,  {Keeton} C.~R.,  2006, \mn@doi [\apj] {10.1086/508769}, \href
  {https://ui.adsabs.harvard.edu/abs/2006ApJ...653.1391D} {653, 1391}

\bibitem[\protect\citeauthoryear{{Falco}, {Gorenstein}  \& {Shapiro}}{{Falco}
  et~al.}{1985}]{1985ApJ...289L...1F}
{Falco} E.~E.,  {Gorenstein} M.~V.,   {Shapiro} I.~I.,  1985, \mn@doi [\apjl]
  {10.1086/184422}, \href
  {https://ui.adsabs.harvard.edu/abs/1985ApJ...289L...1F} {289, L1}

\bibitem[\protect\citeauthoryear{{Foxley-Marrable}, {Collett}, {Vernardos},
  {Goldstein}  \& {Bacon}}{{Foxley-Marrable}
  et~al.}{2018}]{2018MNRAS.478.5081F}
{Foxley-Marrable} M.,  {Collett} T.~E.,  {Vernardos} G.,  {Goldstein} D.~A.,
  {Bacon} D.,  2018, \mn@doi [\mnras] {10.1093/mnras/sty1346}, \href
  {https://ui.adsabs.harvard.edu/abs/2018MNRAS.478.5081F} {478, 5081}

\bibitem[\protect\citeauthoryear{{Goldstein}, {Nugent}, {Kasen}  \&
  {Collett}}{{Goldstein} et~al.}{2018}]{2018ApJ...855...22G}
{Goldstein} D.~A.,  {Nugent} P.~E.,  {Kasen} D.~N.,   {Collett} T.~E.,  2018,
  \mn@doi [\apj] {10.3847/1538-4357/aaa975}, \href
  {https://ui.adsabs.harvard.edu/abs/2018ApJ...855...22G} {855, 22}

\bibitem[\protect\citeauthoryear{{Goldstein}, {Nugent}  \&
  {Goobar}}{{Goldstein} et~al.}{2019}]{2019ApJS..243....6G}
{Goldstein} D.~A.,  {Nugent} P.~E.,   {Goobar} A.,  2019, \mn@doi [\apjs]
  {10.3847/1538-4365/ab1fe0}, \href
  {https://ui.adsabs.harvard.edu/abs/2019ApJS..243....6G} {243, 6}

\bibitem[\protect\citeauthoryear{{Granot}, {Schechter}  \&
  {Wambsganss}}{{Granot} et~al.}{2003}]{2003ApJ...583..575G}
{Granot} J.,  {Schechter} P.~L.,   {Wambsganss} J.,  2003, \mn@doi [\apj]
  {10.1086/345447}, \href
  {https://ui.adsabs.harvard.edu/abs/2003ApJ...583..575G} {583, 575}

\bibitem[\protect\citeauthoryear{{Huber} et~al.,}{{Huber}
  et~al.}{2019}]{2019A&A...631A.161H}
{Huber} S.,  et~al., 2019, \mn@doi [\aap] {10.1051/0004-6361/201935370}, \href
  {https://ui.adsabs.harvard.edu/abs/2019A&A...631A.161H} {631, A161}

\bibitem[\protect\citeauthoryear{{Huber} et~al.,}{{Huber}
  et~al.}{2022}]{2022A&A...658A.157H}
{Huber} S.,  et~al., 2022, \mn@doi [\aap] {10.1051/0004-6361/202141956}, \href
  {https://ui.adsabs.harvard.edu/abs/2022A&A...658A.157H} {658, A157}

\bibitem[\protect\citeauthoryear{{Hyde} \& {Bernardi}}{{Hyde} \&
  {Bernardi}}{2009}]{2009MNRAS.394.1978H}
{Hyde} J.~B.,  {Bernardi} M.,  2009, \mn@doi [\mnras]
  {10.1111/j.1365-2966.2009.14445.x}, \href
  {https://ui.adsabs.harvard.edu/abs/2009MNRAS.394.1978H} {394, 1978}

\bibitem[\protect\citeauthoryear{Jospin, Laga, Boussaid, Buntine  \&
  Bennamoun}{Jospin et~al.}{2022}]{2022BNN}
Jospin L.,  Laga H.,  Boussaid F.,  Buntine W.,   Bennamoun M.,  2022, \mn@doi
  [IEEE Computational Intelligence Magazine] {10.1109/MCI.2022.3155327}, 17, 29

\bibitem[\protect\citeauthoryear{{Katz}, {Balbus}  \& {Paczynski}}{{Katz}
  et~al.}{1986}]{1986ApJ...306....2K}
{Katz} N.,  {Balbus} S.,   {Paczynski} B.,  1986, \mn@doi [\apj]
  {10.1086/164313}, \href
  {https://ui.adsabs.harvard.edu/abs/1986ApJ...306....2K} {306, 2}

\bibitem[\protect\citeauthoryear{{Kayser}, {Refsdal}  \& {Stabell}}{{Kayser}
  et~al.}{1986}]{1986A&A...166...36K}
{Kayser} R.,  {Refsdal} S.,   {Stabell} R.,  1986, \aap, \href
  {https://ui.adsabs.harvard.edu/abs/1986A&A...166...36K} {166, 36}

\bibitem[\protect\citeauthoryear{{Kenworthy} et~al.,}{{Kenworthy}
  et~al.}{2021}]{2021ApJ...923..265K}
{Kenworthy} W.~D.,  et~al., 2021, \mn@doi [\apj] {10.3847/1538-4357/ac30d8},
  \href {https://ui.adsabs.harvard.edu/abs/2021ApJ...923..265K} {923, 265}

\bibitem[\protect\citeauthoryear{{Kochanek}}{{Kochanek}}{2004}]{2004ApJ...605...58K}
{Kochanek} C.~S.,  2004, \mn@doi [\apj] {10.1086/382180}, \href
  {https://ui.adsabs.harvard.edu/abs/2004ApJ...605...58K} {605, 58}

\bibitem[\protect\citeauthoryear{{Lewis} \& {Gil-Merino}}{{Lewis} \&
  {Gil-Merino}}{2006}]{2006ApJ...645..835L}
{Lewis} G.~F.,  {Gil-Merino} R.,  2006, \mn@doi [\apj] {10.1086/504579}, \href
  {https://ui.adsabs.harvard.edu/abs/2006ApJ...645..835L} {645, 835}

\bibitem[\protect\citeauthoryear{{Millon} et~al.,}{{Millon}
  et~al.}{2020}]{2020A&A...640A.105M}
{Millon} M.,  et~al., 2020, \mn@doi [\aap] {10.1051/0004-6361/202037740}, \href
  {https://ui.adsabs.harvard.edu/abs/2020A&A...640A.105M} {640, A105}

\bibitem[\protect\citeauthoryear{{Mortonson}, {Schechter}  \&
  {Wambsganss}}{{Mortonson} et~al.}{2005}]{2005ApJ...628..594M}
{Mortonson} M.~J.,  {Schechter} P.~L.,   {Wambsganss} J.,  2005, \mn@doi [\apj]
  {10.1086/431195}, \href
  {https://ui.adsabs.harvard.edu/abs/2005ApJ...628..594M} {628, 594}

\bibitem[\protect\citeauthoryear{{M{\"o}rtsell}, {Johansson}, {Dhawan},
  {Goobar}, {Amanullah}  \& {Goldstein}}{{M{\"o}rtsell}
  et~al.}{2020}]{2020MNRAS.496.3270M}
{M{\"o}rtsell} E.,  {Johansson} J.,  {Dhawan} S.,  {Goobar} A.,  {Amanullah}
  R.,   {Goldstein} D.~A.,  2020, \mn@doi [\mnras] {10.1093/mnras/staa1600},
  \href {https://ui.adsabs.harvard.edu/abs/2020MNRAS.496.3270M} {496, 3270}

\bibitem[\protect\citeauthoryear{{Narayan}}{{Narayan}}{1991}]{1991ApJ...378L...5N}
{Narayan} R.,  1991, \mn@doi [\apjl] {10.1086/186129}, \href
  {https://ui.adsabs.harvard.edu/abs/1991ApJ...378L...5N} {378, L5}

\bibitem[\protect\citeauthoryear{{Paczynski}}{{Paczynski}}{1986}]{1986ApJ...301..503P}
{Paczynski} B.,  1986, \mn@doi [\apj] {10.1086/163919}, \href
  {https://ui.adsabs.harvard.edu/abs/1986ApJ...301..503P} {301, 503}

\bibitem[\protect\citeauthoryear{{Pan}}{{Pan}}{2020}]{2020ApJ...895L...5P}
{Pan} Y.-C.,  2020, \mn@doi [\apjl] {10.3847/2041-8213/ab8e47}, \href
  {https://ui.adsabs.harvard.edu/abs/2020ApJ...895L...5P} {895, L5}

\bibitem[\protect\citeauthoryear{{Pierel} \& {Rodney}}{{Pierel} \&
  {Rodney}}{2019}]{2019ApJ...876..107P}
{Pierel} J.~D.~R.,  {Rodney} S.,  2019, \mn@doi [\apj]
  {10.3847/1538-4357/ab164a}, \href
  {https://ui.adsabs.harvard.edu/abs/2019ApJ...876..107P} {876, 107}

\bibitem[\protect\citeauthoryear{{Pierel}, {Rodney}, {Vernardos}, {Oguri},
  {Kessler}  \& {Anguita}}{{Pierel} et~al.}{2021}]{2021ApJ...908..190P}
{Pierel} J.~D.~R.,  {Rodney} S.,  {Vernardos} G.,  {Oguri} M.,  {Kessler} R.,
  {Anguita} T.,  2021, \mn@doi [\apj] {10.3847/1538-4357/abd8d3}, \href
  {https://ui.adsabs.harvard.edu/abs/2021ApJ...908..190P} {908, 190}

\bibitem[\protect\citeauthoryear{{Planck Collaboration} et~al.,}{{Planck
  Collaboration} et~al.}{2020}]{2020A&A...641A...6P}
{Planck Collaboration} et~al., 2020, \mn@doi [\aap]
  {10.1051/0004-6361/201833910}, \href
  {https://ui.adsabs.harvard.edu/abs/2020A&A...641A...6P} {641, A6}

\bibitem[\protect\citeauthoryear{{Rauch}, {Mao}, {Wambsganss}  \&
  {Paczynski}}{{Rauch} et~al.}{1992}]{1992ApJ...386...30R}
{Rauch} K.~P.,  {Mao} S.,  {Wambsganss} J.,   {Paczynski} B.,  1992, \mn@doi
  [\apj] {10.1086/170988}, \href
  {https://ui.adsabs.harvard.edu/abs/1992ApJ...386...30R} {386, 30}

\bibitem[\protect\citeauthoryear{{Refsdal}}{{Refsdal}}{1964}]{1964MNRAS.128..307R}
{Refsdal} S.,  1964, \mn@doi [\mnras] {10.1093/mnras/128.4.307}, \href
  {https://ui.adsabs.harvard.edu/abs/1964MNRAS.128..307R} {128, 307}

\bibitem[\protect\citeauthoryear{{Richardson}, {Jenkins}, {Wright}  \&
  {Maddox}}{{Richardson} et~al.}{2014}]{2014AJ....147..118R}
{Richardson} D.,  {Jenkins} Robert~L. I.,  {Wright} J.,   {Maddox} L.,  2014,
  \mn@doi [\aj] {10.1088/0004-6256/147/5/118}, \href
  {https://ui.adsabs.harvard.edu/abs/2014AJ....147..118R} {147, 118}

\bibitem[\protect\citeauthoryear{{Riess} et~al.,}{{Riess}
  et~al.}{2022}]{2022ApJ...934L...7R}
{Riess} A.~G.,  et~al., 2022, \mn@doi [\apjl] {10.3847/2041-8213/ac5c5b}, \href
  {https://ui.adsabs.harvard.edu/abs/2022ApJ...934L...7R} {934, L7}

\bibitem[\protect\citeauthoryear{{Saha} \& {Williams}}{{Saha} \&
  {Williams}}{2011}]{2011MNRAS.411.1671S}
{Saha} P.,  {Williams} L. L.~R.,  2011, \mn@doi [\mnras]
  {10.1111/j.1365-2966.2010.17797.x}, \href
  {https://ui.adsabs.harvard.edu/abs/2011MNRAS.411.1671S} {411, 1671}

\bibitem[\protect\citeauthoryear{{Sainz de Murieta}, {Collett}, {Magee},
  {Weisenbach}, {Krawczyk}  \& {Enzi}}{{Sainz de Murieta}
  et~al.}{2023}]{2023MNRAS.526.4296S}
{Sainz de Murieta} A.,  {Collett} T.~E.,  {Magee} M.~R.,  {Weisenbach} L.,
  {Krawczyk} C.~M.,   {Enzi} W.,  2023, \mn@doi [\mnras]
  {10.1093/mnras/stad3031}, \href
  {https://ui.adsabs.harvard.edu/abs/2023MNRAS.526.4296S} {526, 4296}

\bibitem[\protect\citeauthoryear{{Schechter} \& {Wambsganss}}{{Schechter} \&
  {Wambsganss}}{2002}]{2002ApJ...580..685S}
{Schechter} P.~L.,  {Wambsganss} J.,  2002, \mn@doi [\apj] {10.1086/343856},
  \href {https://ui.adsabs.harvard.edu/abs/2002ApJ...580..685S} {580, 685}

\bibitem[\protect\citeauthoryear{{Schechter}, {Wambsganss}  \&
  {Lewis}}{{Schechter} et~al.}{2004}]{2004ApJ...613...77S}
{Schechter} P.~L.,  {Wambsganss} J.,   {Lewis} G.~F.,  2004, \mn@doi [\apj]
  {10.1086/422907}, \href
  {https://ui.adsabs.harvard.edu/abs/2004ApJ...613...77S} {613, 77}

\bibitem[\protect\citeauthoryear{{Schechter}, {Pooley}, {Blackburne}  \&
  {Wambsganss}}{{Schechter} et~al.}{2014}]{2014ApJ...793...96S}
{Schechter} P.~L.,  {Pooley} D.,  {Blackburne} J.~A.,   {Wambsganss} J.,  2014,
  \mn@doi [\apj] {10.1088/0004-637X/793/2/96}, \href
  {https://ui.adsabs.harvard.edu/abs/2014ApJ...793...96S} {793, 96}

\bibitem[\protect\citeauthoryear{{Schneider} \& {Sluse}}{{Schneider} \&
  {Sluse}}{2014}]{2014A&A...564A.103S}
{Schneider} P.,  {Sluse} D.,  2014, \mn@doi [\aap]
  {10.1051/0004-6361/201322106}, \href
  {https://ui.adsabs.harvard.edu/abs/2014A&A...564A.103S} {564, A103}

\bibitem[\protect\citeauthoryear{{Suyu}, {Goobar}, {Collett}, {More}  \&
  {Vernardos}}{{Suyu} et~al.}{2024}]{2024SSRv..220...13S}
{Suyu} S.~H.,  {Goobar} A.,  {Collett} T.,  {More} A.,   {Vernardos} G.,  2024,
  \mn@doi [\ssr] {10.1007/s11214-024-01044-7}, \href
  {https://ui.adsabs.harvard.edu/abs/2024SSRv..220...13S} {220, 13}

\bibitem[\protect\citeauthoryear{{Treu} \& {Marshall}}{{Treu} \&
  {Marshall}}{2016}]{2016A&ARv..24...11T}
{Treu} T.,  {Marshall} P.~J.,  2016, \mn@doi [\aapr]
  {10.1007/s00159-016-0096-8}, \href
  {https://ui.adsabs.harvard.edu/abs/2016A&ARv..24...11T} {24, 11}

\bibitem[\protect\citeauthoryear{{Vernardos}}{{Vernardos}}{2019}]{2019MNRAS.483.5583V}
{Vernardos} G.,  2019, \mn@doi [\mnras] {10.1093/mnras/sty3486}, \href
  {https://ui.adsabs.harvard.edu/abs/2019MNRAS.483.5583V} {483, 5583}

\bibitem[\protect\citeauthoryear{{Vernardos} \& {Fluke}}{{Vernardos} \&
  {Fluke}}{2013}]{2013MNRAS.434..832V}
{Vernardos} G.,  {Fluke} C.~J.,  2013, \mn@doi [\mnras]
  {10.1093/mnras/stt1076}, \href
  {https://ui.adsabs.harvard.edu/abs/2013MNRAS.434..832V} {434, 832}

\bibitem[\protect\citeauthoryear{{Vernardos} \& {Tsagkatakis}}{{Vernardos} \&
  {Tsagkatakis}}{2019}]{2019MNRAS.486.1944V}
{Vernardos} G.,  {Tsagkatakis} G.,  2019, \mn@doi [\mnras]
  {10.1093/mnras/stz868}, \href
  {https://ui.adsabs.harvard.edu/abs/2019MNRAS.486.1944V} {486, 1944}

\bibitem[\protect\citeauthoryear{{Vernardos}, {Fluke}, {Bate}  \&
  {Croton}}{{Vernardos} et~al.}{2014}]{2014ApJS..211...16V}
{Vernardos} G.,  {Fluke} C.~J.,  {Bate} N.~F.,   {Croton} D.,  2014, \mn@doi
  [\apjs] {10.1088/0067-0049/211/1/16}, \href
  {https://ui.adsabs.harvard.edu/abs/2014ApJS..211...16V} {211, 16}

\bibitem[\protect\citeauthoryear{{Vernardos} et~al.,}{{Vernardos}
  et~al.}{2024}]{2024SSRv..220...14V}
{Vernardos} G.,  et~al., 2024, \mn@doi [\ssr] {10.1007/s11214-024-01043-8},
  \href {https://ui.adsabs.harvard.edu/abs/2024SSRv..220...14V} {220, 14}

\bibitem[\protect\citeauthoryear{{Wambsganss}}{{Wambsganss}}{1990}]{1990PhDT.......180W}
{Wambsganss} J.,  1990, PhD thesis, -

\bibitem[\protect\citeauthoryear{{Wambsganss}, {Witt}  \&
  {Schneider}}{{Wambsganss} et~al.}{1992}]{1992A&A...258..591W}
{Wambsganss} J.,  {Witt} H.~J.,   {Schneider} P.,  1992, \aap, \href
  {https://ui.adsabs.harvard.edu/abs/1992A&A...258..591W} {258, 591}

\bibitem[\protect\citeauthoryear{{Weisenbach}, {Schechter}  \&
  {Pontula}}{{Weisenbach} et~al.}{2021}]{2021ApJ...922...70W}
{Weisenbach} L.,  {Schechter} P.~L.,   {Pontula} S.,  2021, \mn@doi [\apj]
  {10.3847/1538-4357/ac2228}, \href
  {https://ui.adsabs.harvard.edu/abs/2021ApJ...922...70W} {922, 70}

\bibitem[\protect\citeauthoryear{{Witt}}{{Witt}}{1990}]{1990A&A...236..311W}
{Witt} H.~J.,  1990, \aap, \href
  {https://ui.adsabs.harvard.edu/abs/1990A&A...236..311W} {236, 311}

\bibitem[\protect\citeauthoryear{{Wojtak}, {Hjorth}  \& {Gall}}{{Wojtak}
  et~al.}{2019}]{2019MNRAS.487.3342W}
{Wojtak} R.,  {Hjorth} J.,   {Gall} C.,  2019, \mn@doi [\mnras]
  {10.1093/mnras/stz1516}, \href
  {https://ui.adsabs.harvard.edu/abs/2019MNRAS.487.3342W} {487, 3342}

\bibitem[\protect\citeauthoryear{{Wong} et~al.,}{{Wong}
  et~al.}{2020}]{2020MNRAS.498.1420W}
{Wong} K.~C.,  et~al., 2020, \mn@doi [\mnras] {10.1093/mnras/stz3094}, \href
  {https://ui.adsabs.harvard.edu/abs/2020MNRAS.498.1420W} {498, 1420}

\bibitem[\protect\citeauthoryear{{Yahalomi}, {Schechter}  \&
  {Wambsganss}}{{Yahalomi} et~al.}{2017}]{2017arXiv171107919Y}
{Yahalomi} D.~A.,  {Schechter} P.~L.,   {Wambsganss} J.,  2017, \mn@doi [arXiv
  e-prints] {10.48550/arXiv.1711.07919}, \href
  {https://ui.adsabs.harvard.edu/abs/2017arXiv171107919Y} {p. arXiv:1711.07919}

\bibitem[\protect\citeauthoryear{{Young}}{{Young}}{1981}]{1981ApJ...244..756Y}
{Young} P.,  1981, \mn@doi [\apj] {10.1086/158752}, \href
  {https://ui.adsabs.harvard.edu/abs/1981ApJ...244..756Y} {244, 756}

\bibitem[\protect\citeauthoryear{{Zheng}, {Chen}, {Li}  \& {Chen}}{{Zheng}
  et~al.}{2022}]{2022ApJ...931..114Z}
{Zheng} W.,  {Chen} X.,  {Li} G.,   {Chen} H.-Z.,  2022, \mn@doi [\apj]
  {10.3847/1538-4357/ac68ea}, \href
  {https://ui.adsabs.harvard.edu/abs/2022ApJ...931..114Z} {931, 114}

\makeatother
\end{thebibliography}




\appendix

\section{Why macrosaddles cannot be standardized}
\label{app:macrosaddles}

\begin{figure}
    \centering
	\includegraphics[width=\columnwidth]{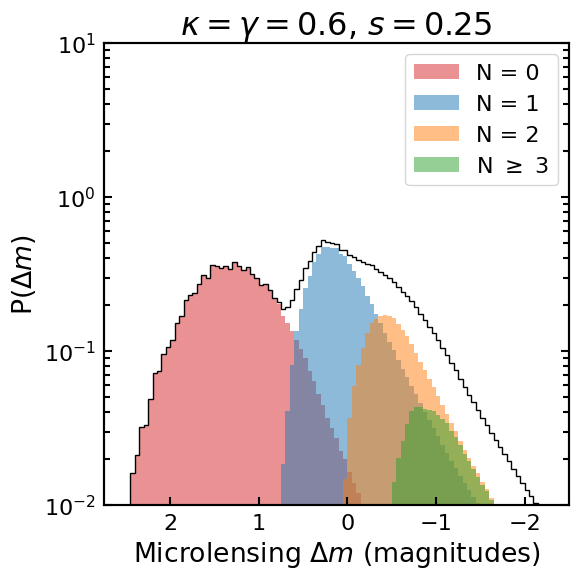}
    \caption{Microlensing magnification histogram for a bright macrosaddle showing the $N=0$ subhistogram.}
    \label{fig:saddlepoint_hist}
\end{figure}

We stated previously that macrosaddles cannot be standardized. It is well known that saddlepoint images are more susceptible to microlensing (de)magnifications \citep{2002ApJ...580..685S}, and the scatter for saddlepoint images can be found in \cite{2021ApJ...922...70W}. The only reason that macrominima can be standardized is that the dominant source of scatter eventually comes from the $N=1$ region. For macrosaddles, there is an $N=0$ region -- i.e. there are \textit{no} microimages which are minima of the time delay surface. Minima are important because they cannot have $\mu < 1$; therefore, the histograms and subhistograms must have hard boundaries for how much they can be demagnified from the macro-model. One might notice in Figure \ref{fig:hists_caustics_zoom} that subhistograms for different values of $N$ take on roughly the same shape; this has been mentioned in the literature but not greatly explored, and we reserve such exploration for future work. Saddlepoint images have no such restriction or lower bound on $\mu$ however, and can become arbitrarily demagnified. Therefore, the subhistograms for $N=0$ regions can look slightly different and have a greater width; see Figure \ref{fig:saddlepoint_hist} or, e.g., Figure 4 from \cite{2003ApJ...583..575G} or Figure 3 from \cite{2011MNRAS.411.1671S}. 

\begin{figure*}
    \centering
	\includegraphics[width=2\columnwidth]{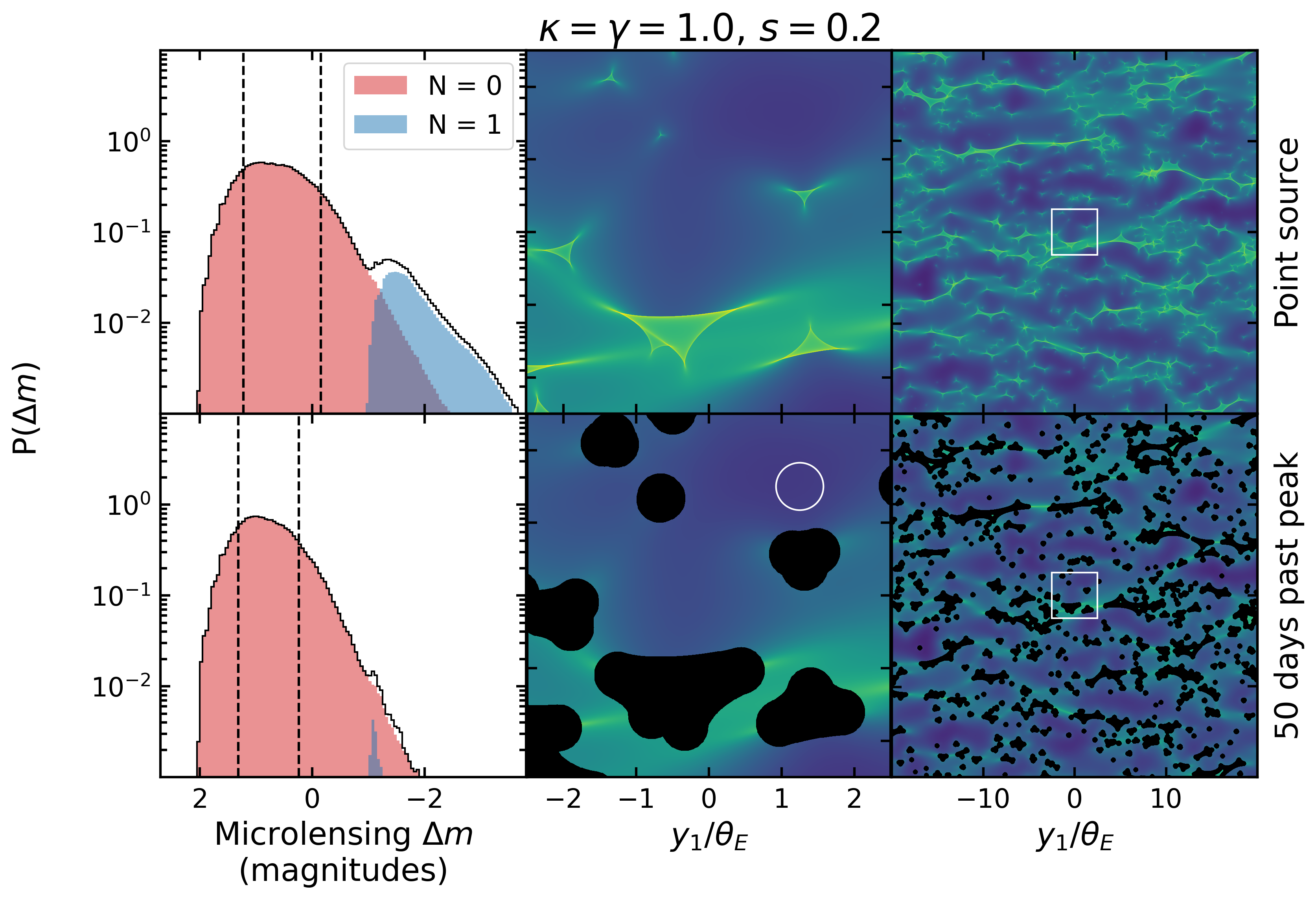}
    \caption{Microlensing magnification histogram for a faint macrosaddle showing the $N=0$ subhistogram. Even when ruling out regions where an expanding supernova would cross a caustic, the low magnification tail of the histogram prevents the system from being standardizable.}
    \label{fig:saddlepoint_hist_2}
\end{figure*}

One might argue that the saddlepoint shown in Figure \ref{fig:saddlepoint_hist} is a bright one with many subhistograms that have similar probabilities, which we have already said makes images unstandardizable. However, the unstandardizability holds true for faint saddlepoints as well, as shown in Figure \ref{fig:saddlepoint_hist_2} -- even if we consider ruling out regions where an expanding supernova would cross a caustic. In essence, we can remove the high magnification tail from minima and saddlepoints, but saddlepoints also have a low magnification tail which cannot be removed. This results in saddlepoint macroimages being unstandardizable. 

\section{SALT3 model errors}

\begin{figure}
    \centering
	\includegraphics[width=\columnwidth]{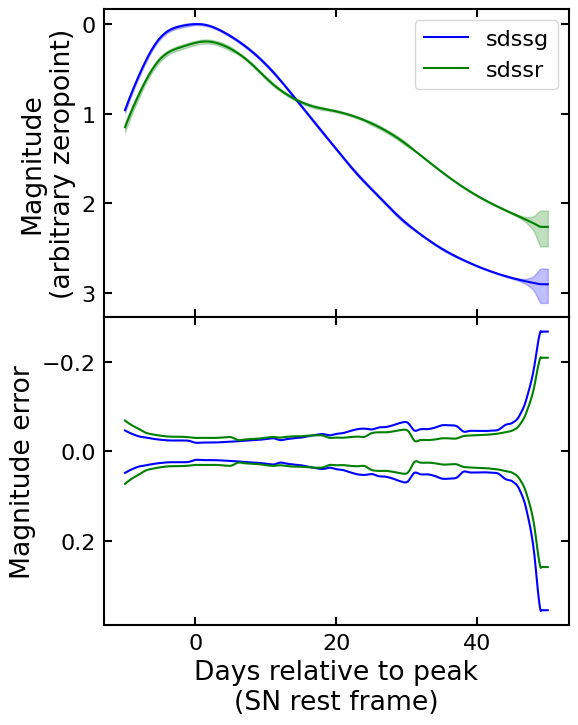}
    \caption{Top: SALT3 supernovae templates with errors due to the model covariances. Bottom: Just the errors as a function of supernova phase.}
    \label{fig:sn_template_errors}
\end{figure}

Figure \ref{fig:sn_template_errors} shows SALT3 supernova templates for the rest frame $g$ and $r$ bands, which will be typically redshifted to the $r$ and $i$ bands in the observer frame. The errors associated with the templates are also shown. For the majority of rest frame phases, the errors are on the order of 0.05 mags. 

\section{Fraction of lightcurves selected}

\begin{figure} 
    \centering
	\includegraphics[width=\columnwidth]{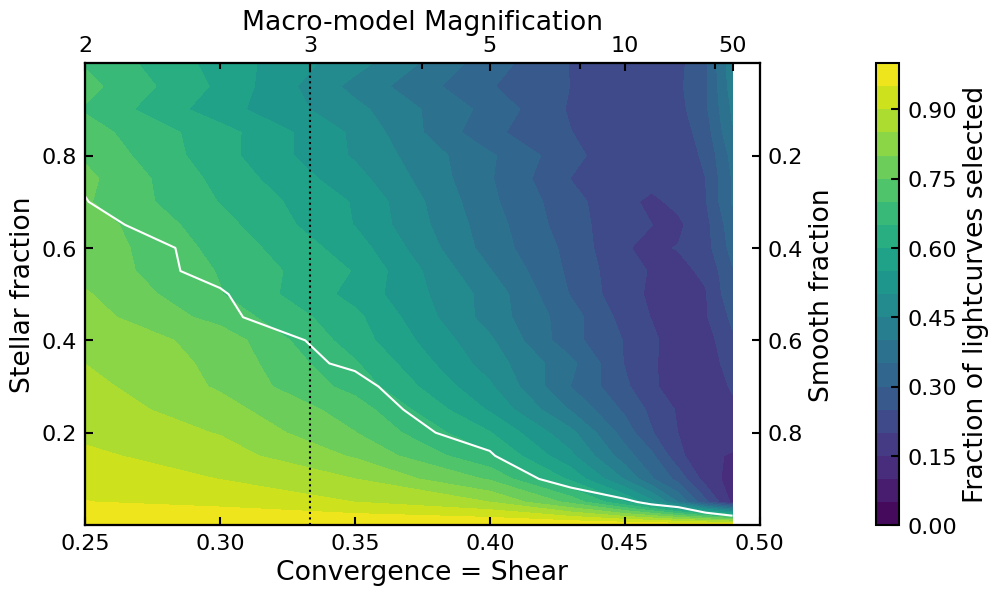}
    \caption{Contour plot showing the fraction of lightcurves which the neural network predicts did not show any caustic crossings. The lightcurves used assume we observe every 2 days from 5 days before peak up to 50 days after peak (in the rest frame of the supernova) and have 0.05 mag noise on the microlensing lightcurve. The solid white line denotes the contour of 0.15 mag microlensing scatter for the lightcurves. The vertical black dotted line marks the boundary where the counter-image of a singular isothermal sphere is demagnified (left) or magnified (right).}
    \label{fig:fraction_selected}
\end{figure}

Figure \ref{fig:fraction_selected} shows the fraction of lightcurves which the neural network is 75\% confident did not have any caustic crossings as a function of lens macro-model parameters. For the majority of standardizable glSNe Ia, at least 70\% of the lightcurves should be useable (flat). 

\section{Propogation of uncertainties}
\label{app:prop_of_uncertainties}

Scaling the mass model by the mass sheet parameter $\lambda$ and introducing the presence of a mass sheet of convergence $\kappa = 1-\lambda$ causes the magnification to transform as \begin{equation}
    \mu \rightarrow \lambda^{-2}\mu.
\end{equation} Given the known intrinsic brightness of a supernovae and its observed brightness, we can infer the magnification. By comparison with the magnification predicted by the model, we can constrain the value of $\lambda$. In the absence of microlensing, a single standardizable glSN Ia can constrain the value of $\lambda$ to within $\approx$ 5\% fractional uncertainty \citep{2020MNRAS.496.3270M, 2022ApJ...924....2B}. 

Using the fact that $\lambda$ is a ratio of model and observed magnifications, \begin{equation}
    \lambda^2 = \frac{\mu_{\text{model}}}{\mu_{\text{obs}}}
\end{equation} and the change in magnitudes from lensing is \begin{equation}
    \Delta m = -2.5\log\mu = \frac{-2.5}{\ln 10} \ln\mu
\end{equation} the fractional uncertainty on the mass sheet parameter $\lambda$ from an individual system is \begin{align}
    \Big(\frac{\sigma_\lambda}{\lambda}\Big)^2 &= \Big(\frac{1}{2}\Big)^2 \Big[\Big(\frac{\sigma_{\text{obs}}}{\mu_{\text{obs}}}\Big)^2 + \Big(\frac{\sigma_{\text{model}}}{\mu_{\text{model}}}\Big)^2\Big]\\
    &= \Big(\frac{\ln 10}{5}\Big)^2\Big(\sigma_{\Delta m_{\text{obs}}}^2 + \sigma_{\Delta m_{\text{model}}}^2\Big)
\end{align} where $\sigma_{\Delta m_{\text{obs}}}$ is the uncertainty on the observed magnification of the supernova and $\sigma_{\Delta m_{\text{model}}}$ is the uncertainty on the magnification of the model, in magnitudes. 


\bsp	
\label{lastpage}
\end{document}